\documentclass[12pt,epsf,epsfig,rotate,graphicx]{article}
\usepackage{graphicx}

\def\simlt{\mathrel{\raise.3ex\hbox{$<$\kern-.75em\lower1ex\hbox{$\sim$}}}}
\def\simgt{\mathrel{\raise.3ex\hbox{$>$\kern-.75em\lower1ex\hbox{$\sim$}}}}

\begin{document}

\begin{titlepage}
\begin{flushright}
IZTECH-P/2005-04
\end{flushright}

\begin{center}
\vspace{1cm}

{\Large \bf LEP Indications for Two Light Higgs Bosons\\ and
U(1)$^{\prime}$ Model}

\vspace{0.5cm}

{\bf Durmu{\c s} A. Demir$^1$, Levent Solmaz$^{1,2}$ and Saime
Solmaz$^{2}$}

\vspace{.8cm}

{\it $^1$ Department of Physics, Izmir Institute of Technology,
IZTECH, TR35430, Izmir, Turkey}

\vspace{0.4cm}

{\it $^2$ Department of Physics, Bal{\i}kesir University,
Bal{\i}kesir, TR10100, Turkey}

\end{center}
\vspace{1cm}

\begin{abstract}
\medskip
Reanalyses of LEP data have shown preference to two light CP-even
Higgs bosons. We discuss implications of such a Higgs boson
spectrum for the minimal supersymmetric model extended by a
Standard Model singlet chiral superfield and an additional Abelian
gauge invariance (the U(1)$^{\prime}$ model). We, in particular,
determine parameter regions that lead to two light CP-even Higgs
bosons while satisfying existing bounds on the mass and mixings of
the extra vector boson. In these parameter regions, the
pseudoscalar Higgs is found to be nearly degenerate in mass with
either the lightest or next-to-lightest Higgs boson. Certain
parameters of the U(1)$^{\prime}$ model such as the effective
$\mu$ parameter are found to be significantly bounded by the LEP
two-light-Higgs signal.

\end{abstract}

\bigskip
\bigskip

\begin{flushleft}
IZTECH-P/2005-04 \\
December 2005
\end{flushleft}

\end{titlepage}

\tableofcontents

\section{Introduction}
Supersymmetric models, in particular, the minimal supersymmetric model (MSSM) have been introduced to solve the
gauge hierarchy problem of the Standard Model (SM). However, the MSSM itself suffers from a naturalness problem
concerning the Higgsino Dirac mass nested in the superpotential of the model. This problem, the $\mu$ problem
\cite{muprob}, has been the main source of motivation for extending the MSSM. The point is to replace $\mu$ by a
chiral superfield whose scalar component develops a vacuum expectation value (VEV) to induce an effective $\mu$
parameter at the desired scale. Up to now, there has been two basic models in this direction: the U(1)$^{\prime}$
models (see the reviews \cite{review}) and next-to-minimal supersymmetry  $i.e.$ NMSSM (see \cite{nmssm}). Both
models have interesting phenomenological implications ranging from rare decays to Higgs phenomenology. In this
work, we are primarily interested in the Higgs sector of the U(1)$^{\prime}$ models. If one is to find an
explanation for why $\mu$ parameter (having no relation to soft-breaking sector of the theory) in the MSSM is
stabilized at the weak scale then such extensions of the MSSM seem to offer a phenomenologically viable pathway.

The U(1)$^{\prime}$ models forbid a bare $\mu$ parameter via the additional Abelian gauge invariance,
U(1)$^{\prime}$ symmetry. The model predicts an additional neutral vector boson, Z$^{\prime}$, which mediates
neutral currents and which mixes with the Z boson of the MSSM. There are continuing collider searches for this
extra Z boson, each leading to certain bounds on its couplings and mass \cite{collider}. There is a host of
constraints originating from different observables \cite{pdg}. The most important and direct ones concern bounds
on Z$^{\prime}$ mass and strength of mixing between Z and Z$^{\prime}$ bosons.

The U(1)$^{\prime}$ models generically predict an additional CP-even Higgs boson which typically weigh near
Z$^{\prime}$. The rest are similar to those in the MSSM in terms of their overall scale and dependencies on the
electroweak Higgs doublets (see \cite{97makalesi} and \cite{e6}, for instance).

The recent reanalysis \cite{postlep} of the LEP data by all four
LEP Collaborations has given an indication for two, rather than
one, light Higgs bosons. Although it is not a clear enough signal
to state the existence of two light Higgs bosons in the bulk of
LEP data, all four LEP experiments see a mild excess near $98\,
{\rm GeV}$ with significance of 2.3 standard deviations to be
contrasted with the second signal seen at $114\, {\rm GeV}$ at 1.7
standard deviations. This two-light-Higgs signal has been
interpreted within the framework of the MSSM in
\cite{kane,drees,plehn}. In the MSSM, if the lightest and
next-to-lightest Higgs bosons are to explain the data the overall
scale of the Higgs sector turns out to be rather close to $M_Z$
(as will be seen, this does not have to be so in U(1)$^{\prime}$
models).

The purpose of this work is to determine the implications of the
LEP two-light-Higgs signal within U(1)$^{\prime}$ models in which
the $\mu$ parameter of the MSSM is dynamically generated. The
paper is organized as follows. In Sec. II below we give a brief
overview of the U(1)$^{\prime}$ model. In Sec. III we discuss
bounds on mass and mixings of the Z$^{\prime}$ boson. In Sec. IV
we discus the LEP two-light-Higgs signal along with its MSSM and
U(1)$^{\prime}$ interpretations. In Sec. V we provide a through
analysis several observables, especially the couplings and masses
of the Higgs bosons, by a scan of the parameter space. In Sec. VI
we conclude.

\section{Overview of the U(1)$^\prime$ Models}
In addition to ones in the SM, there can exist new gauge bosons weighing around a ${\rm TeV}$ provided that they
are sufficiently heavy or weakly coupled to the observed matter. Neutral, color-singlet gauge bosons, the
Z$^{\prime}$ bosons, can arise as low-energy manifestations of GUTs \cite{gut}, strings \cite{string} or
dynamical electroweak breaking \cite{dynamic} theories. In this work we will study a minimal U(1)$^{\prime}$
model (in that it differs from the MSSM only by an additional U(1) invariance and by the presence of a single
MSSM-singlet chiral superfield $S$, to be contrasted with models involving a number of singlets or exotics
\cite{e6,secluded}) described in \cite{97makalesi} without referring to its origin. The model is based on the
gauge group
\begin{equation}
\mbox{SU(3)}_{c}\times \mbox{SU(2)}_{L}\times \mbox{U(1)}_{Y} \times \mbox{U(1)}',
\end{equation}
with gauge couplings $g_3,g_2,g_Y,g_{Y^\prime}$, respectively. The matter content includes the MSSM superfields
and a SM singlet $S$, which are all generically assumed to be charged under the additional U(1)$^\prime$ gauge
symmetry. Explicitly, the particle content is: $\widehat{L}_i \sim (1,2,-1/2, Q_{L})$, $\widehat{E}_i^{c} \sim
(1, 1, 1, Q_{E})$, $\widehat{Q}_i \sim (3, 2, 1/6, Q_{Q})$, $\widehat{U}_i^{c} \sim (\bar{3}, 1, -2/3, Q_{U})$,
$\widehat{D}_i^{c} \sim (\bar{3}, 1, 1/3, Q_{D})$, $\widehat{H}_{d} \sim (1, 2, -1/2, Q_{H_d})$,
$\widehat{H}_{u} \sim (1, 2, 1/2, Q_{H_u})$, $\widehat{S}\sim (1, 1, 0, Q_{S})$, in which $i$ is the family
index.

The superpotential includes a Yukawa coupling of the two electroweak Higgs doublets $H_{u,d}$ to the singlet $S$
as well as the top quark Yukawa coupling:
\begin{equation}
W=h_{s}\widehat{S} \widehat{H}_{u}\cdot\widehat{H}_{d} + h_t \widehat{U}_3^c\widehat{Q}_3\cdot\widehat{H}_u
\label{superpot}
\end{equation}
whose  gauge invariance under U(1)$^\prime$ requires that $Q_{H_u}+Q_{H_d}+ Q_{S} = 0$ and
$Q_{Q_3}+Q_{U_3}+Q_{H_u}=0$. Appearance of a {\it bare} $\mu$ parameter in the superpotential is completely
forbidden as long as $Q_S \neq 0$. In analyzing the model we will always impose this constraint on charges.

In (\ref{superpot}) we have kept only the top quark Yukawa coupling. The neglect of all the light fermion
contributions to the superpotential, especially those of the bottom quark and tau lepton, is justified as long as
we remain in low $\langle H_u\rangle/\langle H_d\rangle\equiv \tan\beta$ domain so that hierarchy of the fermion
masses ($e.g.$ $m_b/m_t$) is generated by the corresponding Yukawa couplings themselves.

As we are primarily interested in the third family, in what follows we shall suppress the family index $i.e.$ we
take $Q_{Q_3} \equiv Q_Q$, $Q_{U_3} \equiv Q_U$ and $Q_{D_3} \equiv Q_D$. The soft-breaking terms relevant for
our analysis are given by
\begin{eqnarray}
\label{soft} -{\cal L}_{soft}&\ni& \left(A_s h_{s} S H_{u}\cdot H_{d}+ A_t h_t \widetilde{U}^c \widetilde{Q}\cdot
H_u+ \mbox{h.c.} \right)\nonumber\\ &+& m_{u}^{2}|H_{u}|^2 +  m_{d}^{2}|H_{d}|^2
+m_{s}^{2}|S|^2+M_{\widetilde{Q}}^2|\widetilde{Q}|^2+ M_{\widetilde{U}}^2|\widetilde{U}|^2 +
M_{\widetilde{D}}^2|\widetilde{D}|^2
\end{eqnarray}
where $A_s$ and $A_t$ are holomorphic trilinear couplings pertaining to Higgs and stop sectors, respectively.
Clearly, there is no reason to expect them to be universal at the weak scale even if they are at the MSSM GUT
scale \cite{97makalesi}. In general, the gaugino masses and soft trilinear couplings $A_{s,t}$ of (\ref{soft})
can be complex; if so, they can provide sources of CP violation (without loss of generality, the Yukawa couplings
$h_{s,t}$ can be assumed to be real). However, for simplicity and definiteness we take all soft parameters real
$i.e.$ we restrict our discussions to CP-conserving theory.

The model at hand provides a dynamical origin for certain parameters in the MSSM Higgs sector. Indeed, below the
scale of U(1)$^{\prime}$ breakdown the $\mu$ parameter of the MSSM is induced to be
\begin{eqnarray}
\mu_{eff} = h_s\, \langle S \rangle \equiv \frac{h_s}{\sqrt{2}}\, v_s
\end{eqnarray}
where the Higgs bilinear soft mass $B$ of the MSSM is given by
\begin{eqnarray}
B_{eff}= \mu_{eff}\, A_s\,.
\end{eqnarray}
These effective parameters suggest that MSSM is an effective theory to be completed by U(1)$^{\prime}$ gauge
invariance with a chiral superfield $S$ above $\langle S \rangle \equiv v_s/\sqrt{2}$.

Going back to the superpotential (\ref{superpot}), the truncation
of the Yukawa sector to top quark Yukawa interaction rests on the
assumption that $\tan\beta$ does not rise to large values.
Notably, in U(1)$^{\prime}$ models $\tan\beta \sim 1$ is not
disfavored if not preferred. (As an example, one recalls that the
'large trilinear coupling minimum' --- the minimum of the
potential that occurs when trilinear couplings are hierarchically
larger than the soft mass-squareds of the Higgs fields --- which
has been extensively studied in \cite{97makalesi,secluded}
exhibits a strong preference to $\tan\beta\simeq 1$. However, this
is no more than an example of existence. In fact, according to
existing bounds Z$^{\prime}$ boson is to weigh well above the Z
boson -- unless certain specific assumptions $e.g.$ leptophobicity
are not made -- and thus this specific minimum is not expected to
arise in our analysis.) In what follows we will take $\tan\beta$
to be close to unity when scanning the parameter space.

The Z and Z$^{\prime}$ bosons acquire their masses by eating, respectively, $\mbox{Im}\left[-\sin\beta H_u^0 +
\cos\beta H_d^0\right]$ and $\mbox{Im}\left[\cos\alpha \cos\beta H_u^0 + \cos\alpha \sin\beta H_d^0-\sin\alpha
S\right]$ where
\begin{eqnarray}
\label{alfa} \cot\alpha =\frac{v}{v_s} \sin\beta \cos\beta
\end{eqnarray}
and $v^2 = v_u^2 + v_d^2$ with $v_u^2/2 \equiv  \langle H_u^0 \rangle^2$,  $v_d^2/2 \equiv \langle H_d^0
\rangle^2$. Clearly, as $v_s \rightarrow \infty$, $\alpha \rightarrow \pi/2$. The remaining neutral degrees of
freedom ${\cal{B}}= \Big\{$ $\mbox{Re}\left[H_u^0\right]-\langle H_u^0 \rangle$,
$\mbox{Re}\left[H_d^0\right]-\langle H_d^0 \rangle$, $\mbox{Re}\left[S\right]-\langle S \rangle$,
$\mbox{Im}\left[\sin\alpha \cos\beta H_u^0 + \sin\alpha \sin\beta H_d^0+\cos\alpha S \right]$ $\Big\}$ span the
space of massive scalars. The physical Higgs bosons are defined by
\begin{eqnarray}
H_i = {\cal{R}}_{i j} {\cal{B}}_{j}
\end{eqnarray}
where the mixing matrix ${\cal{R}}$ necessarily satisfies
${\cal{R}} {\cal{R}}^T =1$, and it has already been computed up to
one loop order in \cite{97makalesi,everett,han,amini}. In the
CP-conserving limit the theory contains three CP-even, one CP-odd,
and a charged Higgs boson. We will name physical CP-even states as
$H_1=h$, $H_2=H$ and $H_3=H^{\prime}$ with $m_h < m_H <
m_H^{\prime}$, and the CP--odd one as $H_4=A$ with mass $m_A$.
Clearly, $m_A^2$ grows with growing $A_s v_s$ yet this tree-level
expectation is modified by radiative corrections. At tree level,
the lightest Higgs boson mass is bounded as
\begin{eqnarray}  \label{mhupper}
m_{h}^2 \leq M_Z^2 \cos^2 2 \beta + \frac{1}{2} h_s^2 v^2 \sin^2 2\beta + g_Y^{\prime\, 2} \left(Q_{H_d} \cos^2
\beta + Q_{H_u} \sin^2 \beta\right)^2 v^2
\end{eqnarray}
where the first term at right-hand side is nothing but the MSSM
bound where the lightest Higgs is lighter than the Z boson at tree
level. The second term is an $F$-term contribution that also
exists in the NMSSM. The last term, the U(1)$^{\prime}$ D-term
contribution, enhances the upper bound in proportion with
$g_{Y}^{\prime\, 2}$. Hence, rather generically, the
U(1)$^{\prime}$ models are the ones admitting largest $m_{h}$ at
tree level. This property is highly advantageous for accommodating
relatively large values of $m_h$ as there is no need to large
radiative corrections. Indeed, for $m_h \sim 114\, {\rm GeV}$, for
instance, one needs sizeable radiative corrections in the MSSM
whereas in U(1)$^{\prime}$ models this is not needed at all
\cite{05makalesi,han}. However, when $m_h$ tends to take smaller
values, tree- and loop-level contributions to $m_h$ must conspire
to generate $m_h$ correctly. Hence, in small $m_h$ regime the most
severely constrained model (among MSSM, NMSSM and U(1)$^{\prime}$
models) turns out to be the U(1)$^{\prime}$ model. In this sense,
one expects LEP two-light-Higgs signal to bound certain parameters
of the U(1)$^{\prime}$ models in a significant way. For example,
when $m_h$ varies in a certain interval $h_s$ is expected to
remain within a certain bound depending on the size of
U(1)$^{\prime}$ D term contribution, as suggested by
(\ref{mhupper}). The discussions in Sec. V will provide a detailed
analysis of the constraints on U(1)$^{\prime}$ models from LEP II
data by taking into account the radiative corrections to the Higgs
sector. In what follows we will base all estimates on one-loop
Higgs boson masses and mixings computed in \cite{everett}. In the
next section we will discuss certain phenomenological bounds on
mass and couplings of the Z$^{\prime}$ boson to determine the
available parameter space.

\section{Constraints from Z-Z$^\prime$ Mixing}

The Z$^{\prime}$ boson couples to neutral currents of MSSM fields with a strength varying with the
U(1)$^{\prime}$ gauge coupling and U(1)$^{\prime}$ charge of fields. Currently, the main constraints on the
existence of a Z$^{\prime}$ boson stem from: ($i$) precision data on neutral current processes, ($ii$)
modifications in Z boson couplings due to its mixing with Z$^{\prime}$ on and off the Z pole, and ($iii$) direct
searches at high energy colliders. Current bounds carry an unavoidable model dependence since a {\rm TeV}--scale
Z$^{\prime}$ can be of various origin \cite{gut,string,dynamic}. When certain model parameters (say,
U(1)$^{\prime}$ gauge coupling and U(1)$^{\prime}$ charges of the fields) are fixed one can derive bounds on the
remaining parameters (say, Z$^{\prime}$ mass). In this section we will discuss implications of bounds on mixing
between Z and Z$^{\prime}$ bosons on U(1)$^{\prime}$ charge assignment and electroweak breaking parameters.

Within U(1)$^{\prime}$ models, the strongest constraints arise
from the non-observation to date of a Z$^{\prime}$, both from
direct searches \cite{collider,carena-tait} and from indirect
precision tests from Z pole, LEP II and neutral weak current data
\cite{cveticlangacker,indirect}. The Z-Z$^{\prime}$ mixing is
described by the mass-squared matrix\footnote{Our description of
Z-Z$^{\prime}$ mixing is at tree level $i.e.$ we do not include
loop corrections to Z and Z$^{\prime}$ masses as well as to their
mixing mass. Moreover, we neglect possible kinetic mixing between
Z and Z$^{\prime}$ \cite{kinetic}.}
\begin{eqnarray}
\label{mzzp}
M_{Z-Z^\prime}=\left(\begin{array}{cc} M_{Z}^{2} & \Delta^{2}\\\\
\Delta^{2} & M_{Z^\prime}^{2}\end{array}\right),
\end{eqnarray}

where

\begin{eqnarray}
M_{Z}^{2}=G^{2} v^2/4\;,\, M_{Z^\prime}^{2}=g_{Y^\prime}^{2}\left(Q_{H_u}^{2}v_u^{2} + Q_{H_d}^{2} v_d^{2} +
Q_s^{2} v_s^{2}\right)\;,\, \Delta^{2}=\frac{1}{2}g_{Y^\prime} G \left(Q_{H_u} v_u^{2} - Q_{H_d} v_d^{2}\right)
\end{eqnarray}
with $G^2 = g_Y^2 +g_2^2= g_2^2/\cos^2\theta_W$. Current bounds imply that the Z- Z$^{\prime}$ mixing angle,
defined by
\begin{eqnarray}
\label{alphaz} \alpha_{Z-Z^\prime} = \frac{1}{2} \arctan\left(\frac{2
\Delta^{2}}{M_{Z^\prime}^{2}-M_{Z}^{2}}\right)\,,
\end{eqnarray}
should not exceed ${\rm few} \times 10^{-3}$ in absolute magnitude.

\begin{table}[tbp]

\begin{center}

\begin{tabular}{|c||c|c|c|c|c|c|}
\hline  & $Q_{H_u}$ & $Q_{H_d}$ & $ Q_S $ & $ Q_Q $ & $ Q_U $ \\ \hline\hline Model I &   -1  &   -1  &   2 &
1/2 &1/2\\\hline
 Model II&   -1 &    -2 &   3 &  0& 1\\
\hline
\end{tabular}
\end{center}
\caption{\label{table2}{ The U(1)$^{\prime}$ charge assignments of the fields in Model I and Model II. }}
\end{table}

Implications of a small $\alpha_{Z-Z^\prime}$ have already been
analyzed previously \cite{cveticlangacker,97makalesi}. One can see
from (\ref{alphaz}) that unless $M_{Z^\prime}\gg M_Z$, the
Z-Z$^\prime$ mixing angle is naturally of ${\cal{O}}(1)$.
Therefore, a small $\left|\alpha_{Z-Z^\prime}\right|$ requires a
cancellation in the mixing term $\Delta^2$ for a given value of
$\tan\beta$. For models in which $M_{Z^\prime}\sim O(M_Z)$, this
cancellation must be nearly exact. However, this tuning is
alleviated when Z$^{\prime}$ mass is near its natural upper limit
of a few ${\rm TeV}$.  Hence, $\tan^2\beta$ must be tuned around
$Q_{H_d}/Q_{H_u}$ with a precision determined by the size of
${\alpha_{Z-Z^\prime}}$ and how heavy Z$^{\prime}$ is.

In general, larger the mass of Z$^{\prime}$ smaller the fine-tuning needed to suppress $\Delta^2$ and less
severe the impact of  phenomenological bounds. For instance, the assumption of leptophobicity does not stand as
a phenomenological necessity for heavy Z$^{\prime}$. Nevertheless, one should keep in mind that heavier the
Z$^{\prime}$ more difficult to stabilize $\mu_{eff}$ if they are governed by the same Higgs sector. A rather
interesting model which overcomes this difficulty was constructed in \cite{secluded}. This model is, however,
beyond the scope of this work.

\begin{figure}[htb]
\begin{center}
\includegraphics[scale=.5]{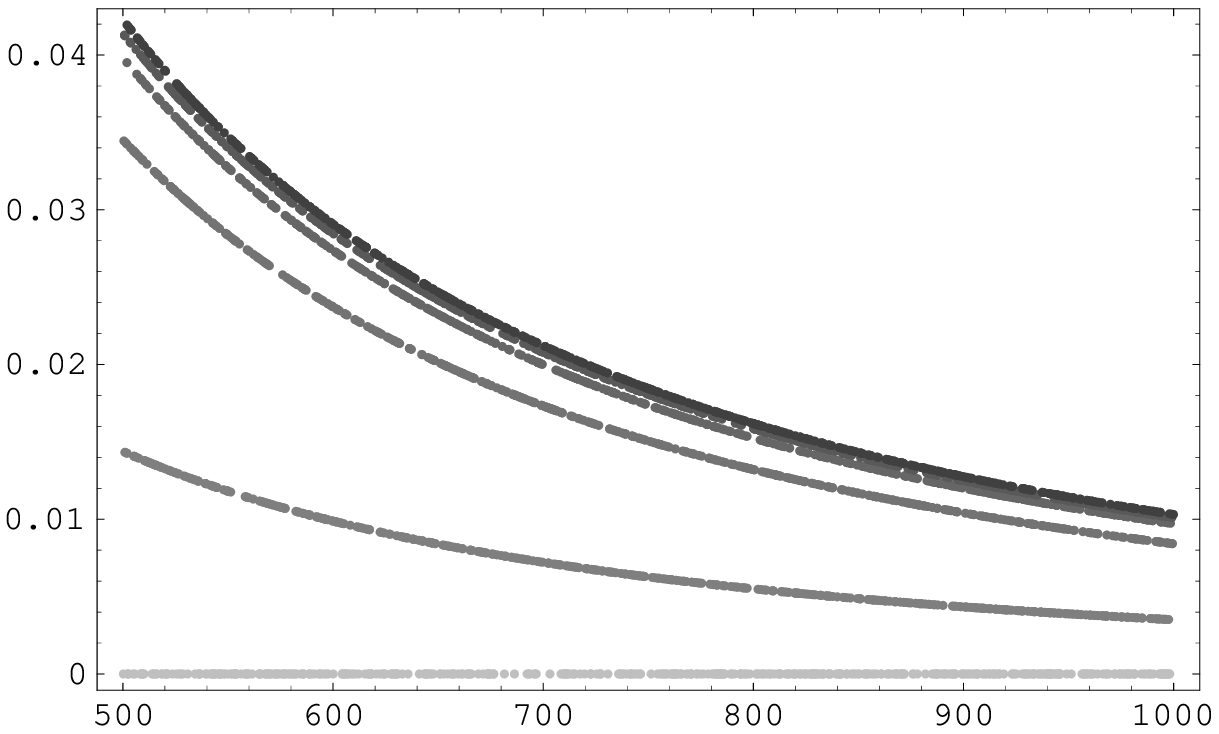} \hspace{1.5cm}
\includegraphics[scale=.5]{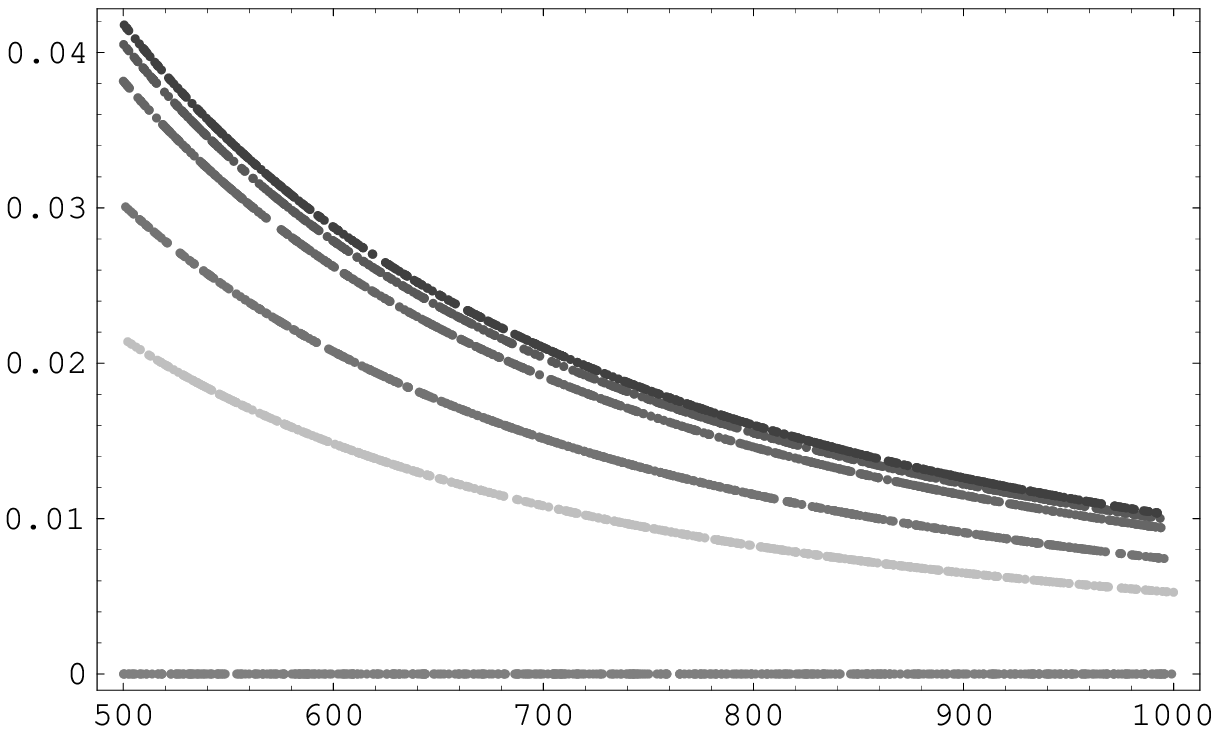} \vspace{0.cm}
\begin{picture}(0,0)(0,0)
     \put(-85,-15){$M_{Z^{\prime}} ({\rm GeV})$}
    \put(-310,-15){$M_{Z^{\prime}} ({\rm GeV})$}
    \put(-435,60){$|\alpha_{Z-Z^{\prime}}|$}
    \put(-210,60){$|\alpha_{Z-Z^{\prime}}|$}
\put(-425,90){$(a)$}
    \put(-205,90){$(b)$}
 \end{picture}
    \end{center}
\caption{\label{quqd0}{ Variation of the Z-Z$^{\prime}$ mixing angle with $M_{Z^{\prime}}$ for different values
of $\tan\beta$. We let $M_{Z^{\prime}}$ vary from $0.5$ to $1\,{\rm TeV}$ and fix U(1)$^{\prime}$ charges of the
Higgs fields as in Model I (panel (a)) and Model II (panel (b)) described in Table \ref{table2}. The shading of
the curves is such that darkness of the curves increases as $\tan\beta$ takes values $\tan\beta
=1$,$\sqrt{2}$,3,5,7 and 10. The brightest curve in panel (a) and the next-to-brightest curve in panel (b)
corresponds, respectively, to $\tan\beta= 1$ and $\sqrt{2}$ for which $\Delta^2$ vanishes exactly.}}
\end{figure}

For definiteness, our numerical analyses will be based on two specific U(1)$^{\prime}$ models, the Model I and
Model II. They are differentiated by the U(1)$^{\prime}$ charge assignments of the fields. For the purpose of
this work, it suffices to fix charges of $H_u$, $H_d$, $S$, $Q$ and $U$, and they are depicted in Table
\ref{table2}. The Model I is taken from \cite{97makalesi} where $Q_{H_u}$ and  $Q_{H_d}$ were chosen to make
$\tan\beta \sim 1$ appropriate for the 'large trilinear coupling vacuum' mentioned in the text. Model II is taken
from a recent discussion \cite{05makalesi} of U(1)$^{\prime}$ models where a family-nonuniversal charge
assignment was used to cancel anomalies of the model such that those fermions whose Yukawa interactions are
forbidden by the family dependence of the charges get their masses from non-holomorphic soft terms, radiatively.
Our discussions here are restricted to holomorphic soft terms with no analysis of anomalies; hence, use of Model
II is, effectively, no more than a specific choice of charges.

In Fig. \ref{quqd0}  we depict the variation of
$\left|\alpha_{Z-Z^{\prime}}\right|$ with $M_{Z^{\prime}}$ for
Model I (left panel) and Model II (right panel). The curves are
for $\tan\beta=1$,$\sqrt{2}$,3,5,7 and 10 whose shadings are
brightest for $\tan\beta=1$ and darkest for $\tan\beta = 10$. One
notices that at $\tan\beta = \sqrt{Q_{H_d}/Q_{H_u}}$ (which equals
to 1 for Model I and $\sqrt{2}$ for model II) the Z-Z$^{\prime}$
mixing angles vanishes exactly irrespective of how heavy
Z$^{\prime}$ is. However, as $\tan\beta$ departs from this
specific value the mixing angle grows rapidly, and it becomes
necessary to increase $M_{Z^{\prime}}$ to higher values to agree
with the bound. Indeed, even for $\tan\beta=\sqrt{2}$ in Model I
(similarly for $\tan\beta = 1$ in Model II) the Z$^{\prime}$ boson
has to weigh $\sim 1.5\, {\rm TeV}$ for
$\left|\alpha_{Z-Z^{\prime}}\right|$ to fall below $10^{-3}$.
Therefore, restriction of $M_{Z^{\prime}}$ below a ${\rm TeV}$
necessarily enforces $\tan\beta$ to remain in close vicinity of
$\sqrt{Q_{H_d}/Q_{H_u}}$. This justifies the truncation  of the
Yukawa sector to top quark couplings in (\ref{superpot}).

This section completes the specification of the U(1)$^{\prime}$ models to be used in the following sections, and
describes the impact of Z-Z$^{\prime}$ mixing angle on model parameters, in particular, on $\tan\beta$. By
examining the response of certain observables to variations in charges (and various soft masses discussed in the
last section) one can trace model-dependence in predictions of the theory. In the next section we will briefly
discuss the LEP two-light-Higgs signal, and its interpretation within the MSSM.

\section{LEP Indications for Two Light Higgs Bosons}

Using $e^+e^-$ collision data at center-of-mass energies between 189 and 209 GeV, the search performed by all
four LEP groups, ALEPH, DELPHI, L3 and OPAL Collaborations, set the lower limit of $114.4\, {\rm  GeV}$ at 95
$\%$ confidence level for the SM Higgs boson \cite{postlep}. Interestingly, in all four experiments there is an
additional common signal of a mild excess near $ 98\, {\rm GeV}$. The signal   around 98 GeV is a 2.3 $\sigma$
effect which should be compared with the 1.7 $\sigma$ excess around 114 GeV. Notably, the former is a weaker
signal than the latter, and if it is not related to background fluctuations or some other experimental
uncertainties then extensions of the SM offering more than one Higgs doublet are favored. Here supersymmetric
models stand as highly viable candidates. In fact, such experimental results can fit quite well to MSSM or its
minimal extensions $i.e.$ NMSSM or U(1)$^{\prime}$ models. Indeed, all these three models have $h$ (the lightest
of all Higgs bosons), $H$ (the next-to-lightest Higgs) and $A$ (the CP--odd Higgs boson) in common. The heavier
Higgs bosons are model-dependent in number and mass range. These Higgs states, if sufficiently light, can
contribute significantly to the formation of four-fermion final states in $e^- e^+$ collisions. In fact,
supersymmetric signals $e^+ e^- \rightarrow (h,H)\, Z$ can give significant contributions especially to two-heavy
fermion signals characterized by final states containing $\overline{b} b\, \overline{f} f$ or $\tau^+ \tau^-\,
\overline{f} f$, $f$ standing for a light fermion. On the other hand, associated production of opposite-CP Higgs
bosons, $e^+ e^- \rightarrow (h, H) A$, can contribute to four-heavy fermion events characterized by final states
consisting of $\overline{b} b\, \overline{f} f$ or $\tau^+ \tau^-\, \overline{f} f$, $f$ standing for $b$ quark
or $\tau$ lepton. Of course, both signals suffer form backgrounds generated by Z boson decays into $\overline{b}
b$ and $\tau^+ \tau^-$.

The MSSM interpretation of the LEP signal \cite{postlep} has already been considered in
\cite{onceki,kane,drees}. The main implication of this two-light-Higgs signal is that the MSSM Higgs sector must
be light as a whole $i.e.$ it should not enter the decoupling regime where $m_H \sim m_A \gg m_h$. In fact, as
has been emphasized in \cite{drees}, the main idea is to identify the signal at $98\, {\rm GeV}$ with $h$ and
the one at $114\, {\rm GeV}$ with $H$. This identification is justified as long as $h Z Z$ coupling is
sufficiently suppressed to cause a relatively weak signal at $98\, {\rm GeV}$. This indeed happens if the
overall mass scale of the Higgs sector is close to $M_Z$. In \cite{kane} discussions were given of various MSSM
parameter regions, including finite CP--odd phases, predicting light Higgs bosons in the LEP data. This analysis
suggests that the requisite range of the $\mu$ parameter is typically ${\cal{O}}(2\, {\rm TeV})$ unless
$m_A\simeq m_h$ within a few ${\rm GeV}$. In general, the relative phase between $A_t$ and $\mu$ provides an
additional freedom for achieving the correct configuration. It is interesting that, according to \cite{kane},
the least fine-tuned parameter space corresponds to a light Higgs boson of mass $m_h \simeq 114\, {\rm GeV}$
with all the rest being heavy. (Here fine-tuning refers to sensitivity of a given parameter set to changes in
parameter values specified at the GUT scale.)

The recent work \cite{plehn} provides a detailed analysis of the two-light-Higgs signal within the CP-conserving
MSSM by imposing bounds from $B_d \rightarrow X_s \gamma$, muon $g-2$, $B_s \rightarrow \mu \overline{\mu}$ as
well as from relic density of the lightest neutralino. The allowed parameter space turns out to be particularly
wide for $\mu \simgt 1\, {\rm TeV}$. The bounds from these observables are found to constrain the MSSM parameter
space unless model parameters are tuned to evade them \cite{plehn}. Consequently, in both CP-conserving
\cite{drees,plehn} and CP-violating \cite{kane} cases the MSSM offers a wide parameter region which provides an
explanation for the LEP two-light-Higgs signal.

In this work we will discuss possible implications of the LEP two-light-Higgs  signal for the U(1)$^{\prime}$
models specified by charge assignments in Table \ref{table2}. Our analyses are based on the radiatively-corrected
Higgs boson masses and mixings computed in \cite{everett}. For the model under concern to explain the data, the
signal strengths must be reproduced correctly at the indicated Higgs mass values. The contribution of
Z$^{\prime}$ mediation is negligible within its mass range, and thus, we focus on the Z boson mediated Higgs
production processes. The Higgs production cross sections depend on all the parameters in the Higgs mass-squared
matrix via the Higgs boson couplings to Z as well as the Higgs boson masses. Leaving their tensor structures
aside, the Higgs-Z-Z couplings are given by \cite{pakile}
\begin{eqnarray}
\label{higgsZ} C_{h Z Z} &=& {\cal{R}}^{h d} \cos \beta + {\cal{R}}^{ h u} \sin\beta\nonumber\\ C_{H Z Z} &=&
{\cal{R}}^{H d} \cos \beta
+ {\cal{R}}^{ H u} \sin\beta\nonumber\\
C_{H^{\prime} Z Z} &=& {\cal{R}}^{H^{\prime} d} \cos \beta
+ {\cal{R}}^{ H^{\prime} u} \sin\beta
\end{eqnarray}
in units of the SM $h Z Z$ coupling $G M_Z$. On the other hand,
coupling of the opposite-CP Higgs bosons to Z are given by
\begin{eqnarray}
C_{h A Z} &=& \sin\alpha\left( {\cal{R}}^{h u} \cos \beta  - {\cal{R}}^{ h d} \sin\beta\right)\nonumber\\
C_{H A Z} &=& \sin\alpha\left( {\cal{R}}^{H u} \cos\beta - {\cal{R}}^{ H d} \sin\beta \right)\nonumber\\
C_{H^{\prime} A Z} &=& \sin\alpha\left({\cal{R}}^{H^{\prime} u} \cos \beta - {\cal{R}}^{ H^{\prime} d}
\sin\beta\right)
\end{eqnarray}
in units of $G/2$, where $G=\sqrt{g_Y^2+g_2^2}$ as defined before. Here ${\cal{R}}$ is the Higgs mixing matrix
defined in Sec. II. The notation is such that ${\cal{R}}^{h d}$, for instance, denotes the entry of ${\cal{R}}$
formed by the row corresponding to lightest Higgs boson $h$ and by the column corresponding to the neutral
CP-even component, $\phi_d$, of $H_d$. The Higgs mass matrix is taken in the basis ${\cal{B}}$ given in Sec. II.

These couplings govern what Higgs bosons are produced with what strength if they are kinematically accessible.
The number of excess events around $98\, {\rm GeV}$ forms about  $10 \%$ of the events which would be generated
by the SM Higgs boson production with $m_{h_{SM}} = 98\, {\rm GeV}$. More quantitatively, the cross sections
satisfy
\begin{eqnarray}
\frac{\sigma(e^+ e^- \rightarrow h Z)}{\sigma(e^+ e^- \rightarrow h_{SM} Z)} = C_{h Z Z}^2 \simeq 0.1
\end{eqnarray}
if $m_h =m_{h_{SM}} = 98\, {\rm GeV}$. Hence, given the statistical significances of the two signals at $98$ and
$114\, {\rm GeV}$, the parameter ranges favored by the LEP excess events turn out to be
\begin{eqnarray}
\label{bounds} 95\, {\rm GeV} \leq m_h \leq 101\, {\rm GeV}\,,\;\, 111\, {\rm GeV} \leq m_H \leq 119\, {\rm
GeV}\,,\;\, 0.056 \leq C_{h Z Z}^2 \leq  0.144
\end{eqnarray}
as has first been derived by \cite{drees} while analyzing the signal within the MSSM. The strength of the $114\,
{\rm GeV}$ signal, with respect to the SM expectation, depends on the coupling strength of $H^{\prime}$ to the Z
boson: $C_{H Z Z}^2 \simeq  0.9 - C_{H^{\prime} Z Z}^2$. However, when Z$^{\prime}$ is heavy so is $H^{\prime}$
and $C_{H Z Z}^2$ turns out to be rather close to the MSSM expectation. In the opposite limit $i.e.$ when
Z$^{\prime}$ weighs relatively light so is $H^{\prime}$, and $C_{H^{\prime} Z Z}^2$ becomes too large to allow
$C_{H Z Z}^2$ to remain close to its MSSM counterpart. These parameter domains will be illustrated by scanning
the parameter space in the next section.

Clearly, $(h, H, H^{\prime})$-Z-Z and $(h, H, H^{\prime})$-$A$-Z couplings are correlated with each other. The
strength of correlation depends on how light $H^{\prime}$ is, that is, how close U(1)$^{\prime}$ breaking scale
is to $M_Z$. For instance, for heavy $H^{\prime}$ the singlet components of the remaining Higgs bosons are
suppressed, $\alpha \rightarrow \pi/2$, and one finds $C_{H Z Z}^2 \simeq C_{h A Z}^2 \simeq 0.9$. This enhances
the $h A$ production compared to $H A$ production, if they are kinematically accessible. Nevertheless, one keeps
in mind that productions of opposite-CP Higgs bosons are P-wave suppressed; moreover, LEP data have not yet been
subjected to a global analysis like \cite{postlep} for such final states.

In the next section we will provide a scan of the U(1)$^{\prime}$ parameter space to determine allowed regions
and correlations among the model parameters under the LEP constraints (\ref{bounds}).

\section{Confronting U(1)$^{\prime}$ Models with LEP Data}
\label{facelep}
 In this section we will determine constraints on the parameters of U(1)$^{\prime}$
model from the LEP two-light-Higgs signal. Before imposing the LEP bounds (\ref{bounds}), we list down allowed
ranges or values of the model parameters. These choices, which stem from different reasons, bring considerable
ease in scanning of the parameter space:
\begin{itemize}
\item $M_{Z^{\prime}} \in \left[0.5, 1\right]\, {\rm TeV}$. This range for $M_{Z^{\prime}}$ is chosen to agree
with bounds from direct collider searches \cite{collider} on one hand, and to prevent $M_{Z^{\prime}}$ slipping
into deep ${\rm TeV}$ domain, on the other hand. The latter introduces a hierarchy problem within the gauge boson
sector \cite{dreesx,secluded}.

\item $|\alpha_{Z-Z^\prime}|\leq 2\times 10^{-3}$. Using this bound together with the aforementioned interval for
$M_{Z^{\prime}}$, $\tan\beta$ is found to remain in close vicinity of $\sqrt{Q_{H_d}/Q_{H_u}}$:  $0.94 \leq
\tan\beta \leq 1.06$ for Model I and $1.36 \leq \tan\beta \leq 1.47$ for Model II.

\item $g_{Y^\prime}^{2}=\frac{5}{3} G^2 \sin^{2}\theta_{W}$. This choice for $g_{Y^\prime}$ might be inspired
from one-step GUT breaking; however, care should be payed to the normalization of the U(1)$^{\prime}$ charges.
Indeed, overall normalization of the charges (as in GUTs, for instance) results in a rescaling of $g_{Y^\prime}$
so that the value quoted here does not need to be the correct choice for U(1)$^{\prime}$ charges in Table
\ref{table2}. Therefore, this equality for $g_{Y^\prime}$ should be regarded as a specific choice, not
necessarily stemming from the GUTs.

\item $h_s\in \left[0.1,0.7\right]$. The RGE studies in \cite{97makalesi,secluded} suggest that $h_s \simlt
 {\cal{O}}(0.7)$ for perturbativity up to the MSSM gauge coupling unification scale.

\item U(1)$^{\prime}$ charges of the fields as in Table \ref{table2}.

\item $M_{\widetilde{Q}},M_{\widetilde{U}}\in \left[ 0.5,5\right] v$, $0< A_{t,s}\simlt 10 v$ and
$m_{\widetilde{t}_1} \geq 100\, {\rm GeV}$, $\widetilde{t}_1$ being the lighter stop. These choices appropriately
put soft-breaking parameters within ${\rm TeV}$ range.
\end{itemize}

In what follows we will impose the LEP bounds (\ref{bounds}) on this parameter space to determine allowed ranges
for model parameters. This determination, depending on how tight it is, will facilitate construction of a
low-energy softly-broken supersymmetric theory devoid of the $\mu$ problem.
\begin{figure}[htb]
\begin{center}
    \includegraphics[scale=.5]{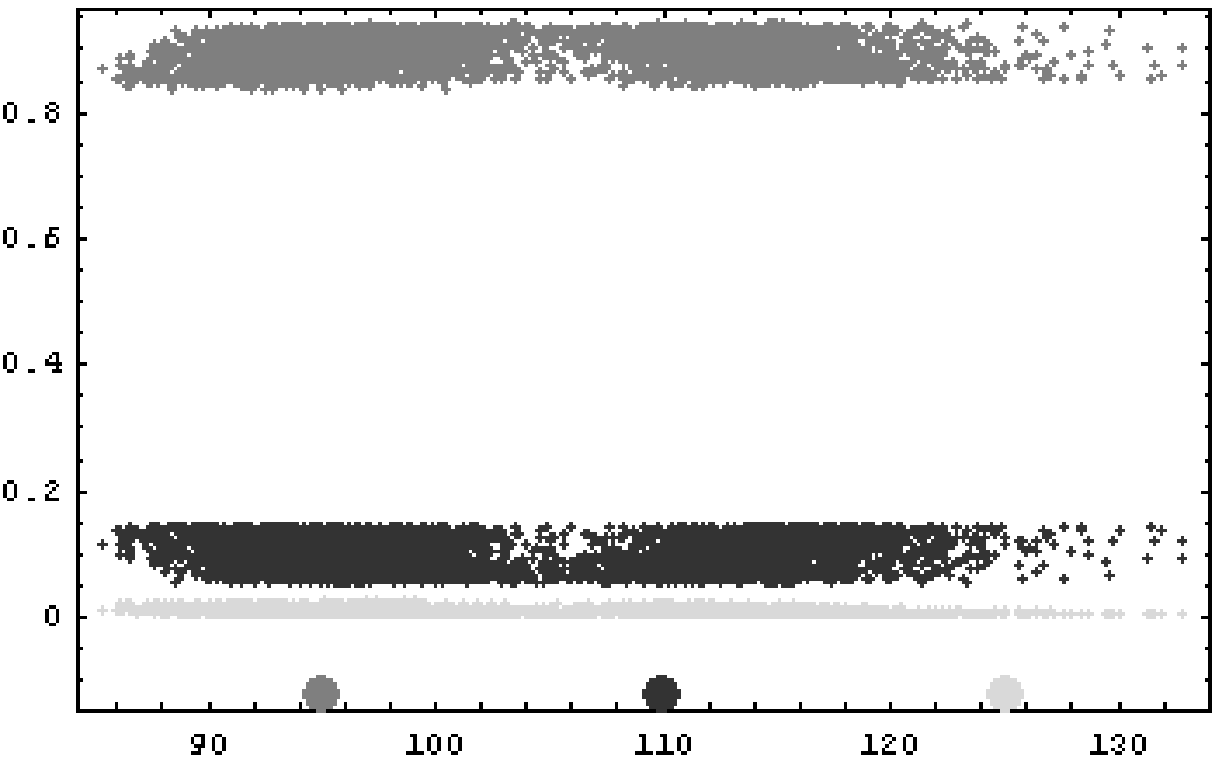}\hspace{1.5cm}
    \includegraphics[scale=.5]{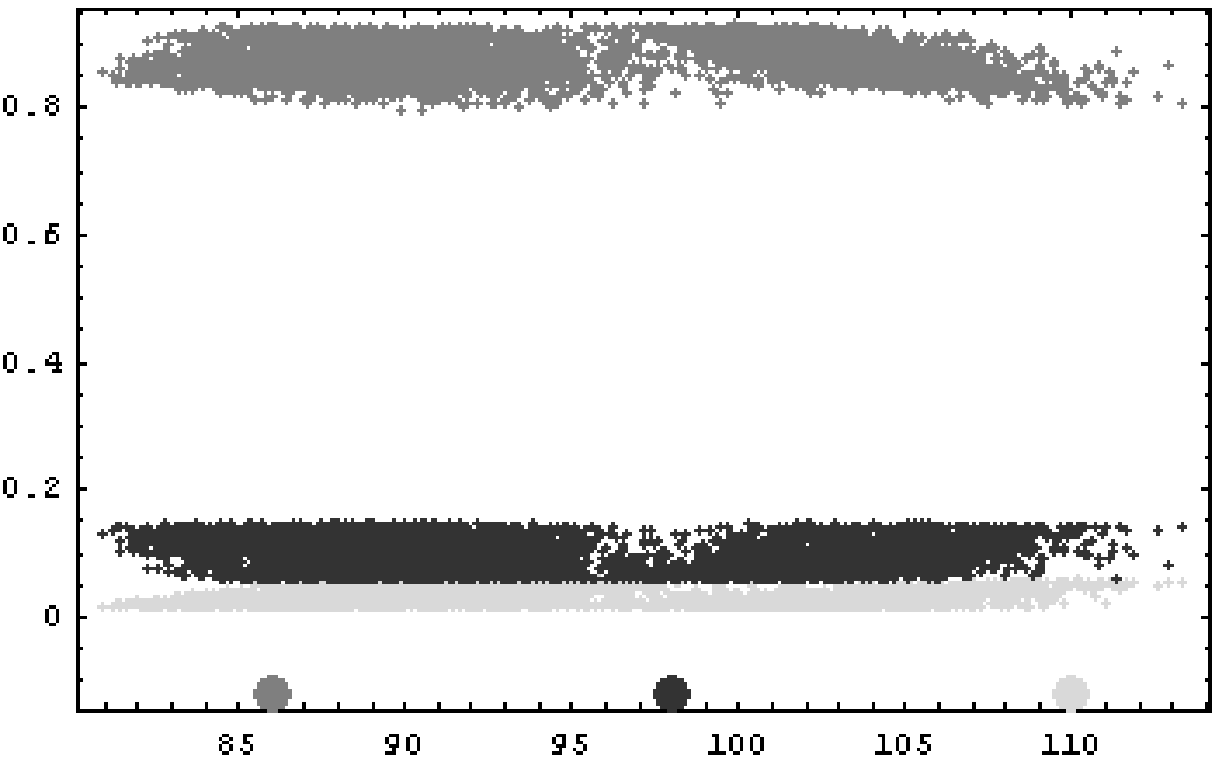}\hspace{0.cm}
\begin{picture}(0,0)(0,0)
     \put(-95,-15){$m_A\,[GeV]$}
    \put(-320,-15){$m_A\,[GeV]$}
    \put(-425,60){$C_{iZZ}^2$}
    \put(-205,60){$C_{iZZ}^2$}
    \put(-425,90){$(a)$}
    \put(-205,90){$(b)$}
          \put(-381,11){\tiny{$H$}}
          \put(-331,11){\tiny{$h$}}
          \put(-284,11){\tiny{${H^\prime}$}}
          \put(-165,11){\tiny{$H$}}
          \put(-110,11){\tiny{$h$}}
          \put(-57,11){\tiny{${H^\prime}$}}
 \end{picture}
    \end{center}
    \caption{\label{cler}{Variation of $C_{(h, H, H^{\prime})ZZ}^2$ with $m_A$ for
Model I (panel (a)) and Model II (panel (b)) after imposing the LEP bounds (\ref{bounds}). Obviously,
$C_{H^{\prime}ZZ}^2$ is rather small (though it can take slightly larger values in Model II than in Model I) and
therefore $C_{H Z Z}^2 \approx 1- C_{h Z Z}^2 \approx 0.9$. The U(1)$^{\prime}$ charge assignments influence
shape of the allowed domains of $C_{(h, H, H^{\prime})ZZ}^2$ as well as their allowed ranges. These figures are
also useful for determining the allowed range of $m_A$: $133 \geq m_A \geq 86$ GeV in Model I and $113\geq m_A
\geq 81$ GeV in Model II. }}
\end{figure}

\begin{figure}[htb]
\begin{center}
    \includegraphics[scale=.5]{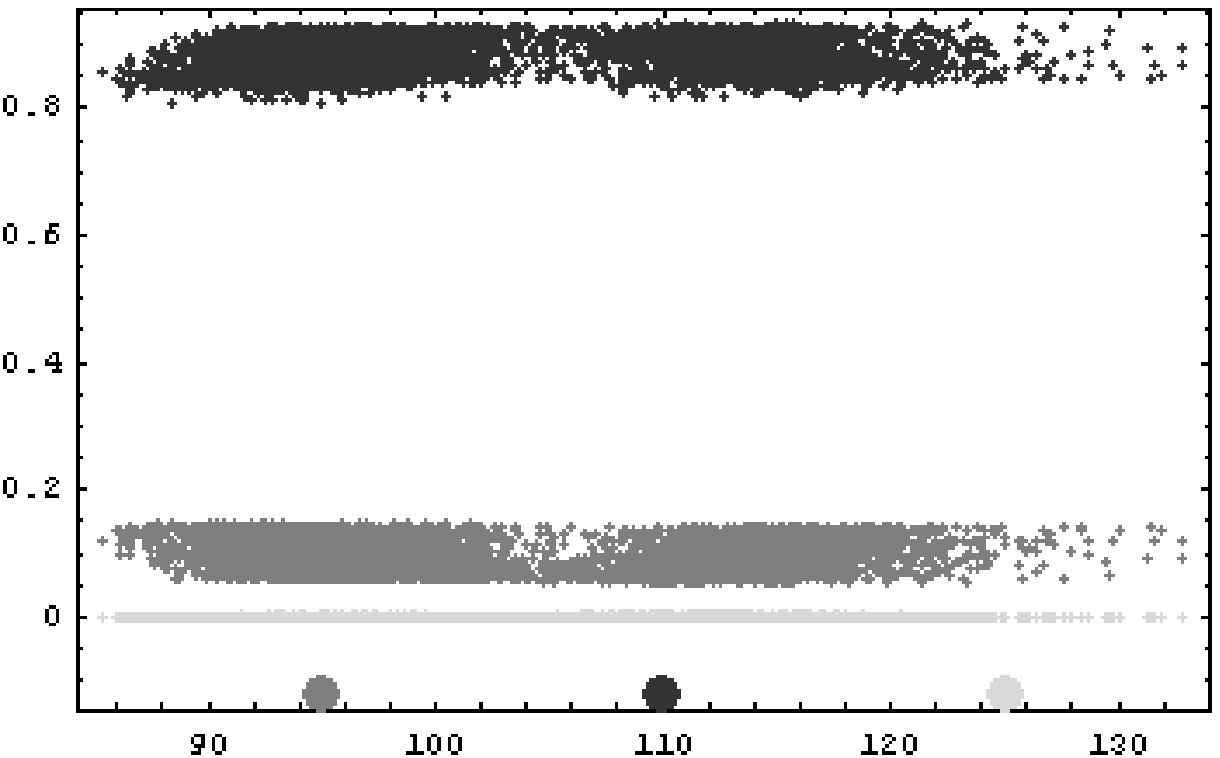}\hspace{1.5cm}
    \includegraphics[scale=.5]{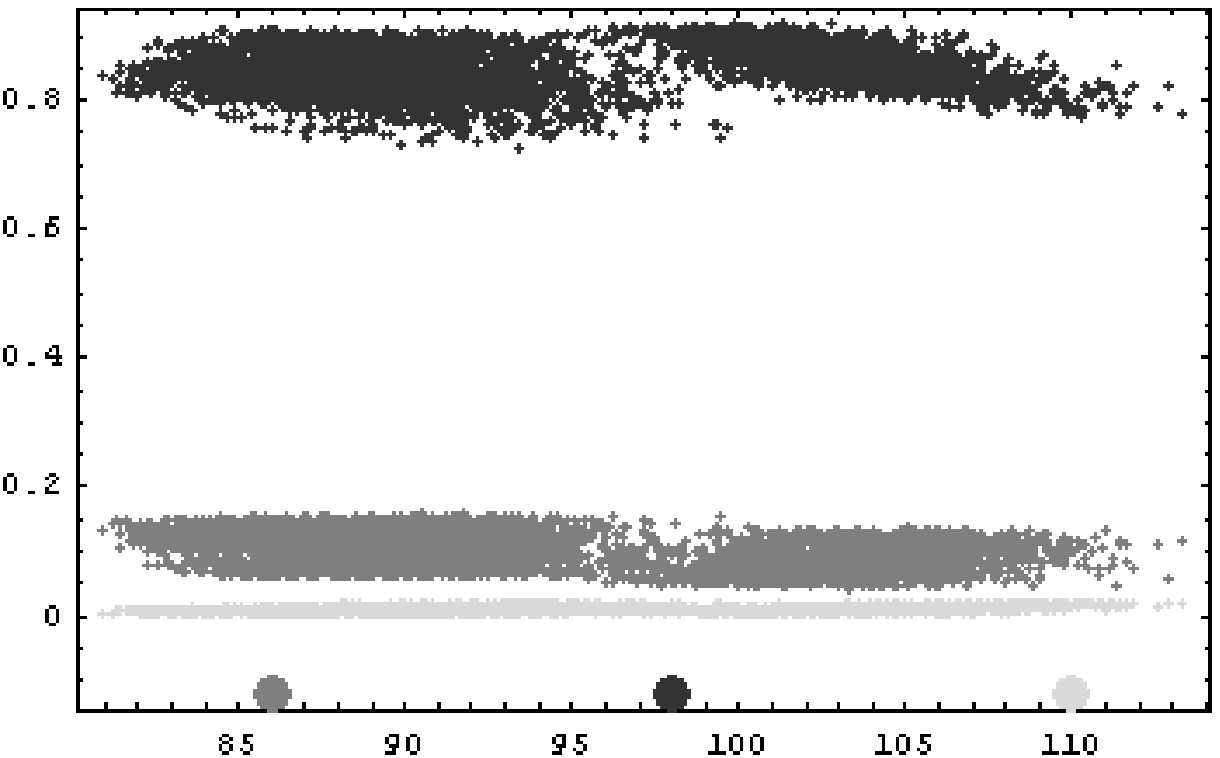}\hspace{0.cm}
\begin{picture}(0,0)(0,0)
     \put(-95,-15){$m_A\,[GeV]$}
    \put(-320,-15){$m_A\,[GeV]$}
    \put(-425,60){$C_{iAZ}^2$}
    \put(-205,60){$C_{iAZ}^2$}
    \put(-425,90){$(a)$}
    \put(-205,90){$(b)$}
      \put(-384,11){\tiny{$H$}}
          \put(-330,11){\tiny{$h$}}
          \put(-281,11){\tiny{${H^\prime}$}}
          \put(-164,11){\tiny{$H$}}
          \put(-109,11){\tiny{$h$}}
          \put(-54,11){\tiny{${H^\prime}$}}
 \end{picture}
    \end{center}
    \caption{\label{cler2}{ Variation of $C_{(h, H, H^{\prime})A Z}^2$ with $m_A$ for
Model I (panel (a)) and Model II (panel (b)) after imposing the LEP bounds (\ref{bounds}). A comparison with Fig.
\ref{cler} reveals that $C_{h A Z}^2 \simeq C_{H Z Z}^2$, $C_{H A Z}^2 \simeq C_{h Z Z}^2$ and $C_{H^{\prime} A
Z}^2 \simeq C_{H^{\prime} Z Z}^2$ as expected from discussions in Sec. IV.}}
\end{figure}

We start the analysis by plotting various Higgs-Z coupling-squareds with respect to the pseudoscalar Higgs mass
$m_A$ by applying the LEP bounds in (\ref{bounds}). $C_{(h, H, H^{\prime})Z Z}^2$ are shown in Fig. \ref{cler}
and $C_{(h, H, H^{\prime})A Z}^2$ in Fig. \ref{cler2} (the shading of each figure is described by the inset in
the panels). This analysis proves useful for determining the (experimentally unconstrained) range of $m_A$.
Indeed, as suggested by the figures, $133 \geq m_A \geq 86$ GeV in Model I and $113\geq m_A \geq 81$ GeV in Model
II. These figures enable one to determine the correlations among various Higgs-Z couplings. First of all,
$C_{H^{\prime} Z Z}^2\ll 1$ and $C_{H^{\prime} Z Z}^2\simlt C_{h Z Z}^2$ for all parameter values of interest.
Therefore, $C_{H Z Z}^2 \simeq  0.9 - C_{H^{\prime} Z Z}^2 \approx  0.9$ as was discussed in Sec. IV.
Furthermore, as comparison of Fig. \ref{cler} and \ref{cler2} reveals, $C_{h A Z}^2 \simeq C_{H Z Z}^2$, $C_{H A
Z}^2 \simeq C_{h Z Z}^2$ and $C_{H^{\prime} A Z}^2 \simeq C_{H^{\prime} Z Z}^2$. Clearly, these correlations
among the couplings become precise when $M_{Z^{\prime}} \rightarrow {\rm TeV}$ since in this case $H^{\prime}$ is
too heavy to have an appreciable doublet component. In the opposite limit $i.e.$ when $M_{Z^{\prime}}$ lies close
to its lower limit, $C_{H^{\prime} Z Z}^2$ can compete with $C_{h Z Z}^2$ so that correlations among the
couplings become too imprecise to compare directly with the MSSM predictions \cite{drees,plehn}.

A comparative look at Figs. \ref{cler} and \ref{cler2} reveals the impact of U(1)$^{\prime}$ charges on Higgs-Z
couplings. Indeed, as U(1)$^{\prime}$ charges are switched from Model I to those of Model II the shapes and
ranges of the allowed domains of couplings change. Obviously, in both models there exist parameter regions where
$C_{H^{\prime} Z Z}^2$ become comparable to $C_{h Z Z}^2$. These effects come in no surprise since, as suggested
by Z-Z$^{\prime}$ mixing, $M_{Z^{\prime}}$ and the Higgs mass-squared matrix, charge assignments influence
various observables. A related point concerns the range of $v_s$. Indeed, for keeping $M_{Z^{\prime}}$ within
$\left[0.5, 1\right] {\rm TeV}$ interval in both models it is necessary to adjust the range of $v_s$ in accord
with the U(1)$^{\prime}$ charges of Higgs fields in the model employed.

\begin{figure}[htb]
\begin{center}
    \includegraphics[scale=.5]{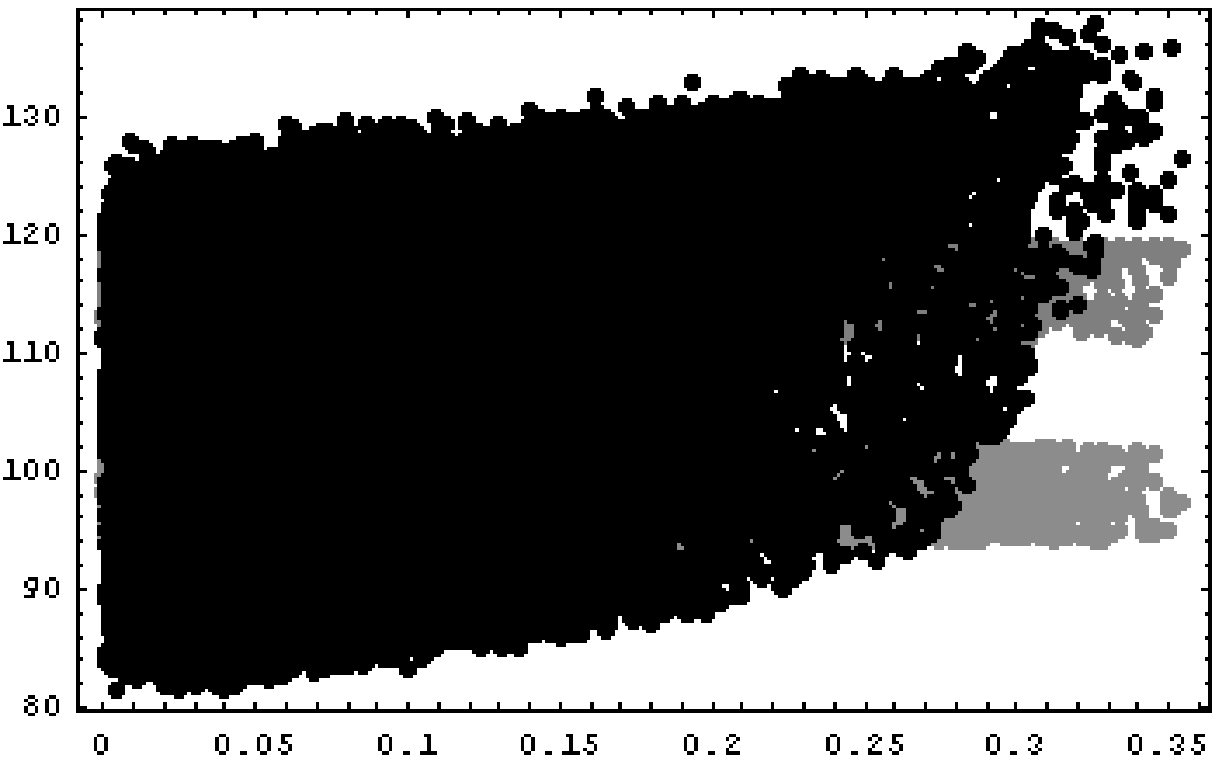}\hspace{1.5cm}
    \includegraphics[scale=.5]{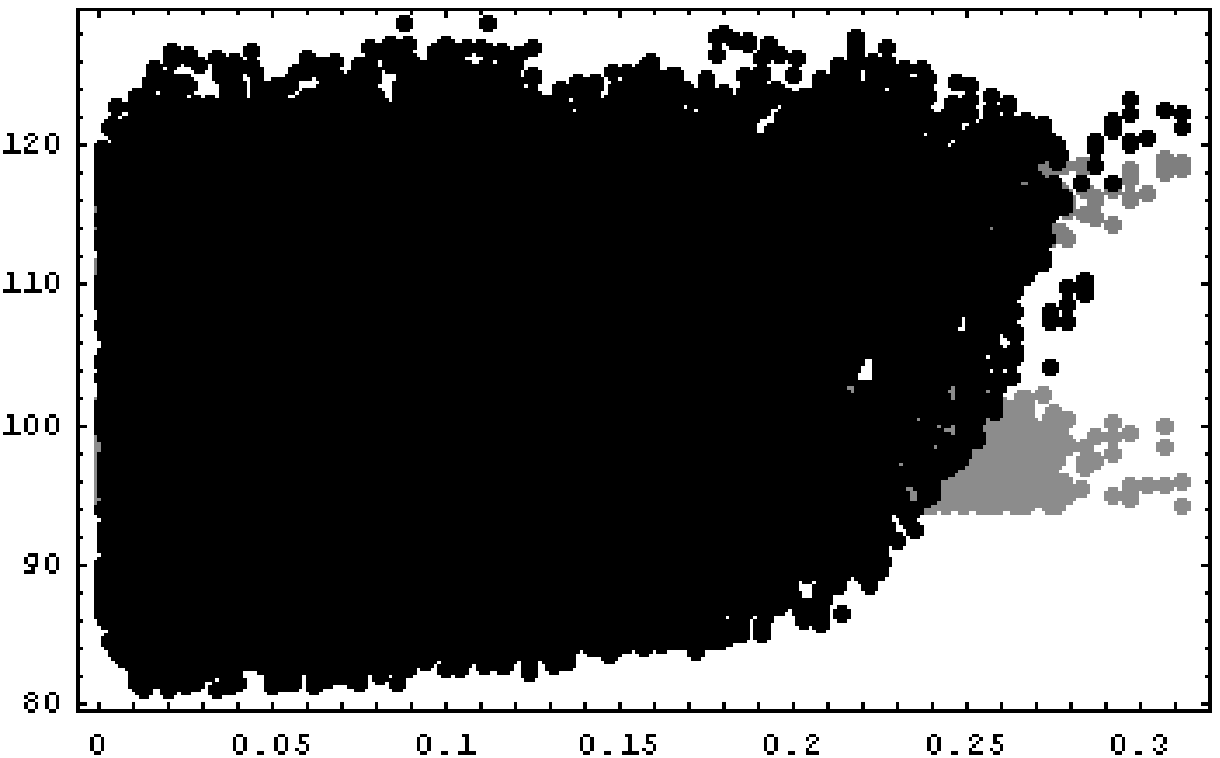}\vspace{1cm}
      \includegraphics[scale=.5]{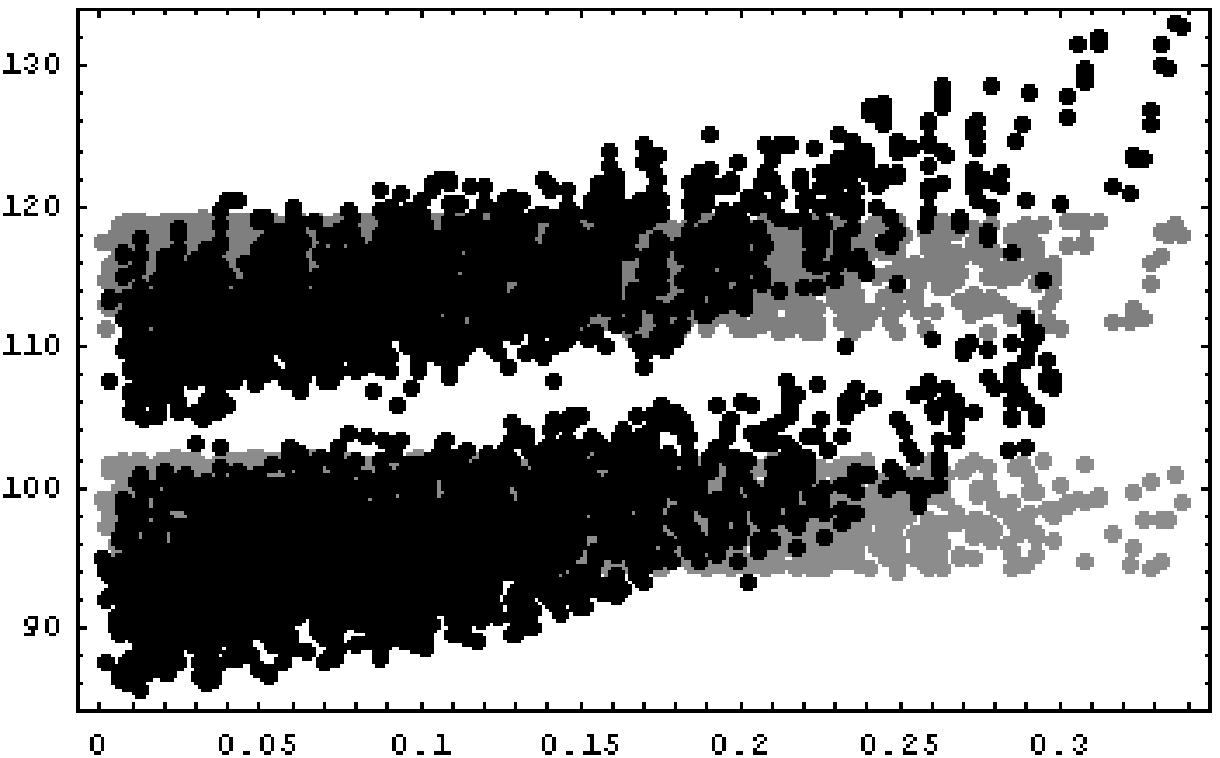}\hspace{1.5cm}
       \includegraphics[scale=.5]{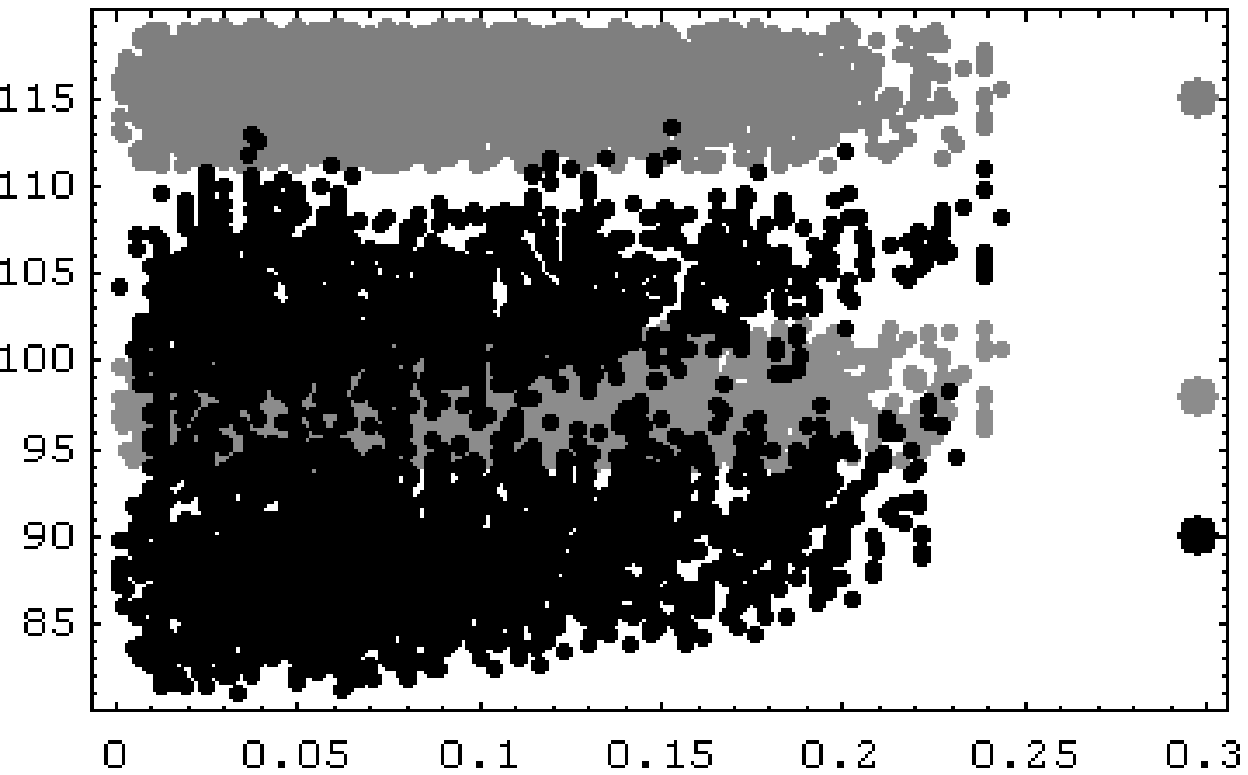}\vspace{.1cm}
\begin{picture}(0,0)(0,0)
      \put(-85,-15){$A_s/v$}
     \put(-85,120){$A_s/v$}
     \put(-205,60){$M_{H}$}
     \put(-205,190){$M_{H}$}
     \put(-310,-15){$A_s/v$}
     \put(-310,120){$A_s/v$}
    \put(-425,60){$M_{H}$}
    \put(-425,190){$M_{H}$}
  \put(-425,220){$(a)$}
    \put(-205,220){$(b)$}
    \put(-425,90){$(c)$}
    \put(-205,90){$(d)$}
\put(-425,180){\tiny{[GeV]}}
 \put(-425,50){\tiny{[GeV]}}
 \put(-205,180){\tiny{[GeV]}}
\put(-205,50){\tiny{[GeV]}}
  \put(-30,53){\tiny{$m_h$}}
          \put(-30,96){\tiny{$m_H$}}
          \put(-30,33){\tiny{$m_A$}}
 \end{picture}
    \end{center}
    \caption{\label{cler2before}{ The impact of LEP bounds on the allowed parameter regions. Depicted are
    variations of Higgs boson masses (whose shadings are defined by inset in panel (d)) with $A_s/v$ in Model I (panels (a) and (c)) and Model II (panels (b) and
    (d)). The Higgs boson masses in panels (a) and (b) are obtained only when the mass constraints
    $m_h \simeq 98\, {\rm GeV}$ and $m_H \simeq 114\, {\rm GeV}$ are taken into account. In these panels the
    pseudoscalar mass $m_A$ is seen to take values in a rather wide range. The panels (c) and (d) illustrate
    impact of the constraint that the signal at $98\, {\rm GeV}$ forms only $\simeq 10 \%$ of the total. This
    constraint, $C_{h Z Z}^2 \simeq 0.1$, is seen to have a significant effect on the allowed ranges of $m_A$. In
    particular, one notes how $m_A$ approximately splits into $m_A^{high}$ (close to $m_H$) and $m_A^{low}$
    (close to $m_h$) domains. Clearly, these two split regions in which $m_A$ could take values vary model to
    model in shape and separation.}}
\end{figure}

\begin{figure}[htb]
\begin{center}
     \includegraphics[scale=.5]{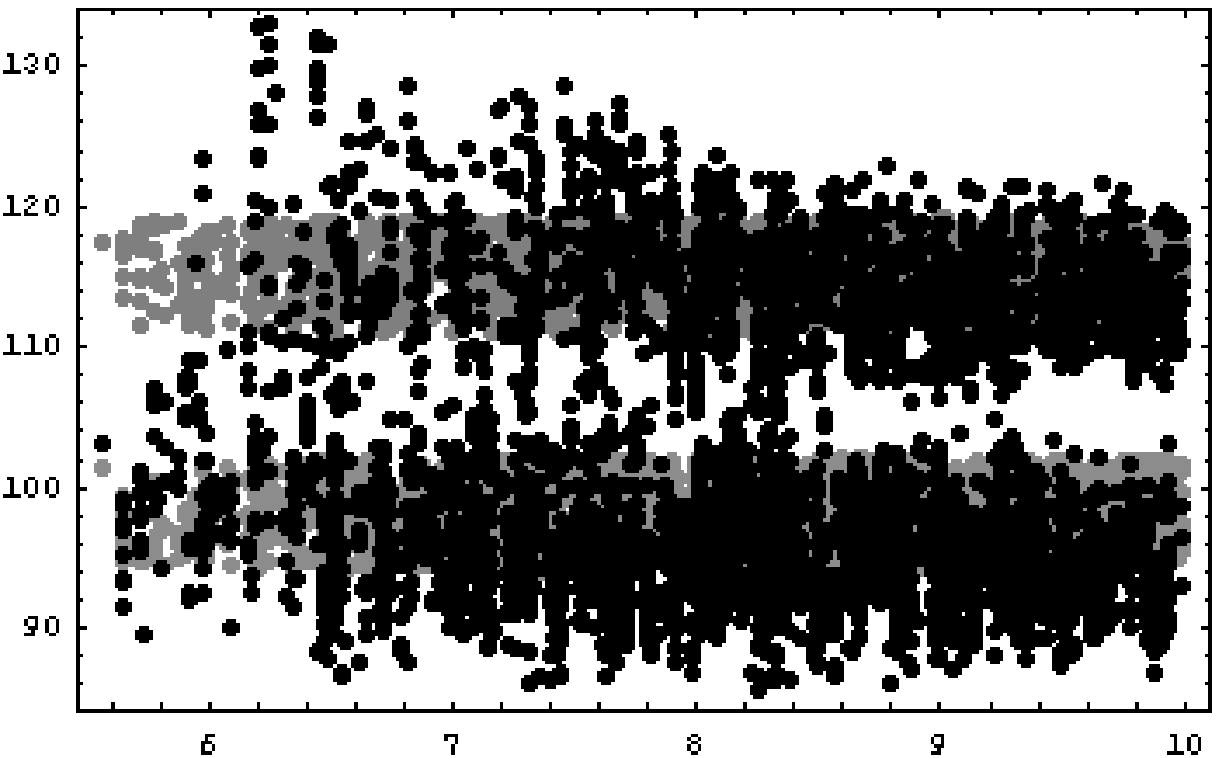} \hspace{1.5cm}
     \includegraphics[scale=.5]{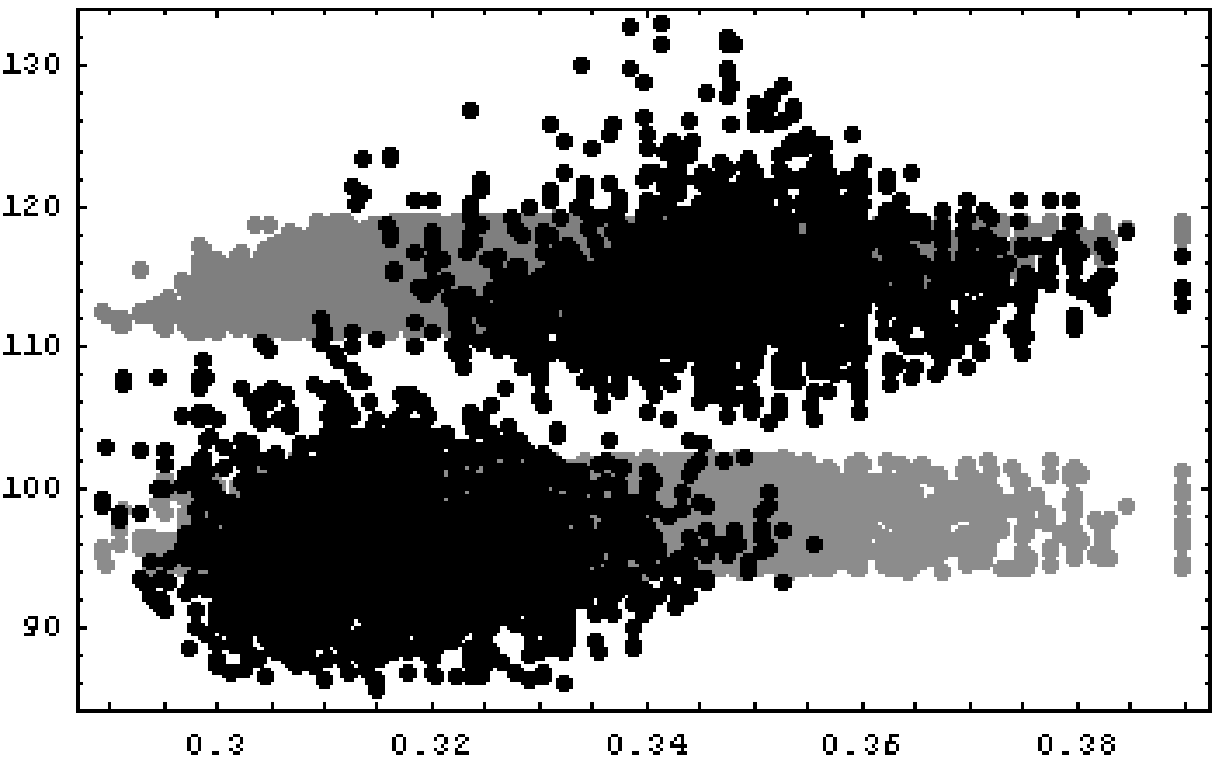}\vspace{1cm}
     \includegraphics[scale=.5]{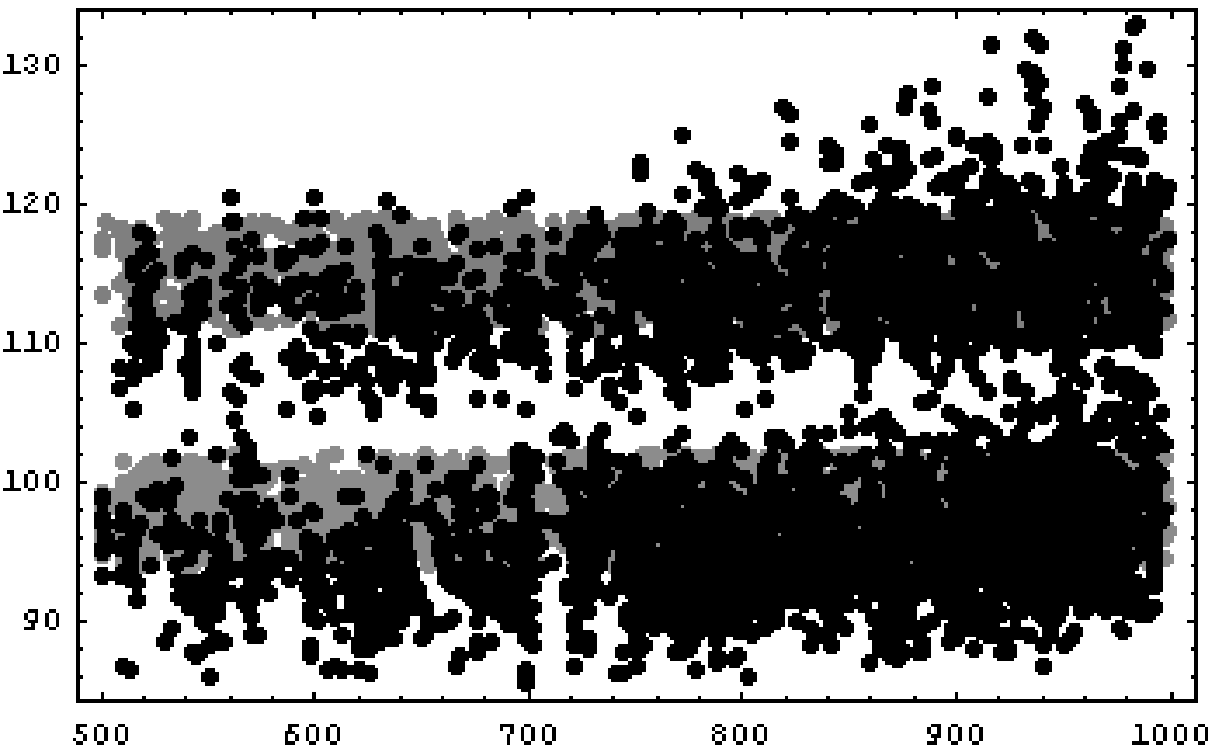}\hspace{1.5cm}
     \includegraphics[scale=.5]{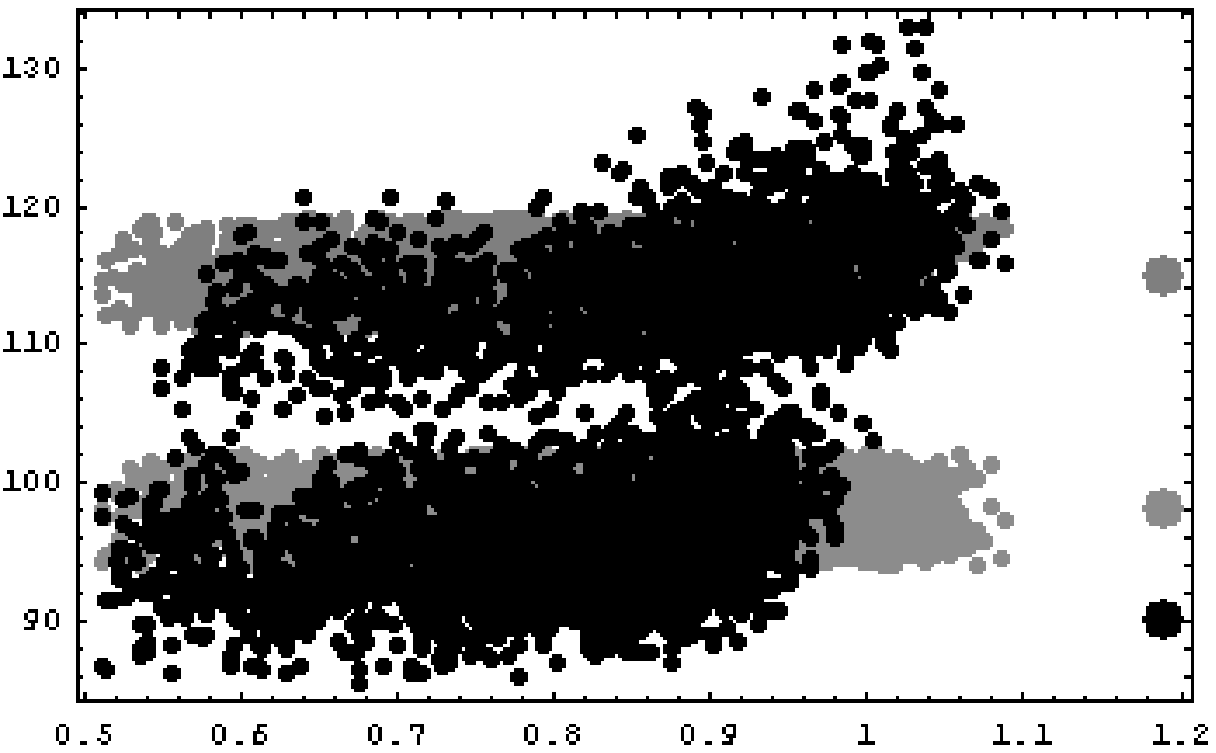}\vspace{.1cm}
\begin{picture}(0,0)(0,0)
   \put(-85,-15){$\mu_{eff}/v$}
     \put(-85,120){$h_s$}
     \put(-205,60){$M_{H}$}
     \put(-205,190){$M_{H}$}
     \put(-310,-15){$M_{Z^\prime}$}
     \put(-310,120){$A_t/v$}
    \put(-425,60){$M_{H}$}
    \put(-425,190){$M_{H}$}
  \put(-425,220){$(a)$}
    \put(-205,220){$(b)$}
    \put(-425,90){$(c)$}
    \put(-205,90){$(d)$}
\put(-425,180){\tiny{[GeV]}} \put(-425,50){\tiny{[GeV]}} \put(-205,180){\tiny{[GeV]}}
\put(-205,50){\tiny{[GeV]}}
     \put(-285,-14){\tiny{[GeV]}}
        \put(-29,35){\tiny{$m_h$}}
          \put(-29,68){\tiny{$m_H$}}
          \put(-29,18){\tiny{$m_A$}}
 \end{picture}
    \end{center}
    \caption{\label{c3m1}{ Variations of the Higgs boson masses  with various parameters in Model I. The CP-even Higgs boson $H^{\prime}$
    is typically degenerate with Z$^{\prime}$ boson, and its mass is not plotted here. (The inset in panel (d) shows
    grey levels used for different Higgs boson masses, as in Fig. \ref{cler2before}.)}}
\end{figure}

\begin{figure}[htb]
\begin{center}
     \includegraphics[scale=.5]{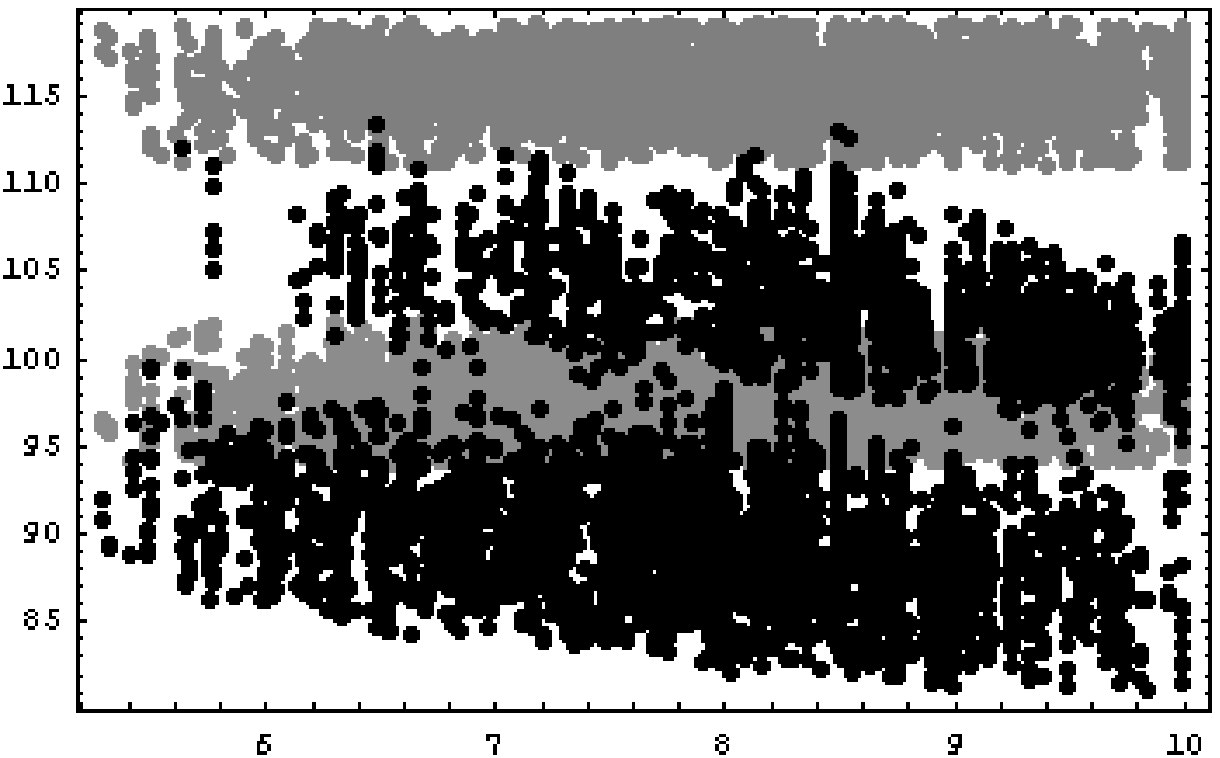} \hspace{1.5cm}
     \includegraphics[scale=.5]{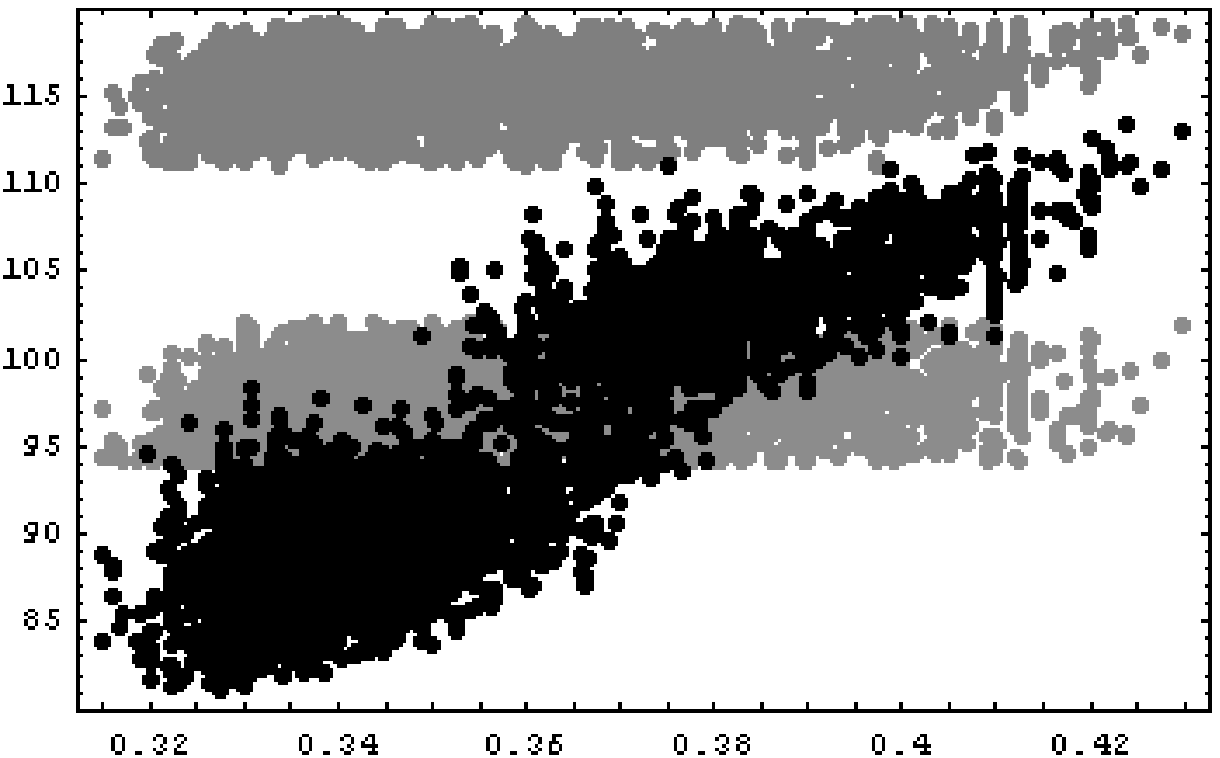}\vspace{1cm}
     \includegraphics[scale=.5]{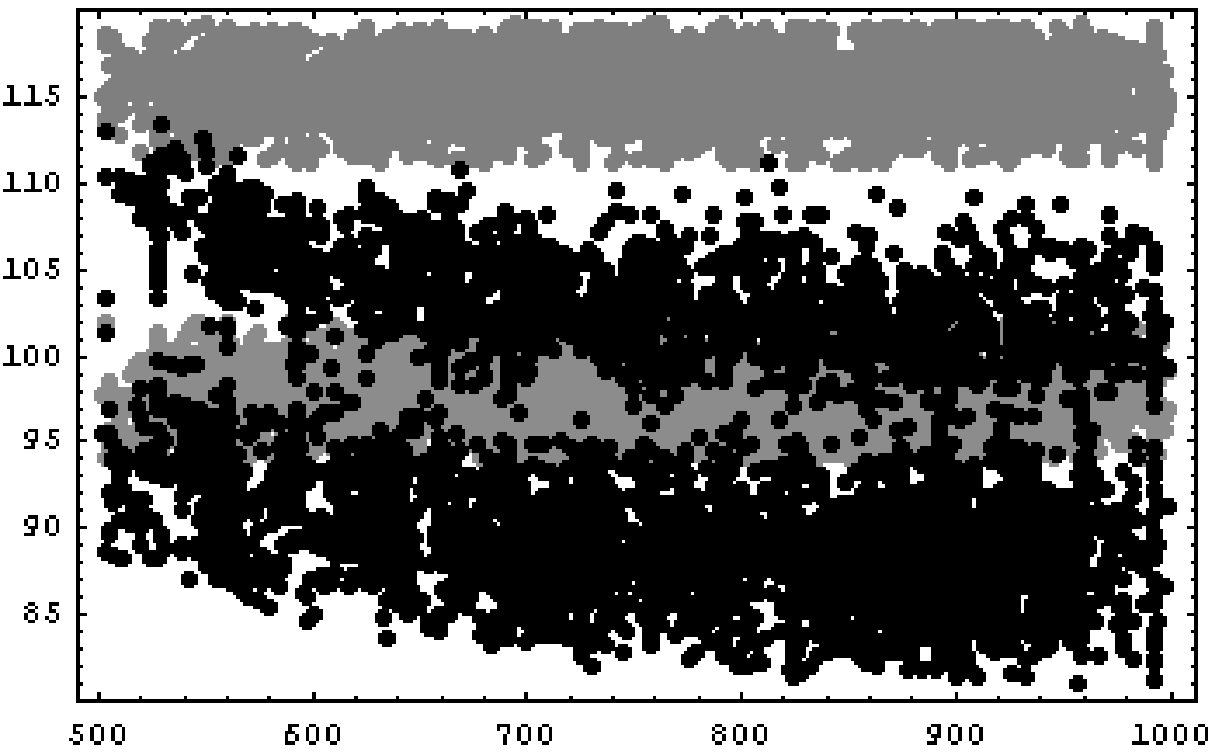}\hspace{1.5cm}
     \includegraphics[scale=.5]{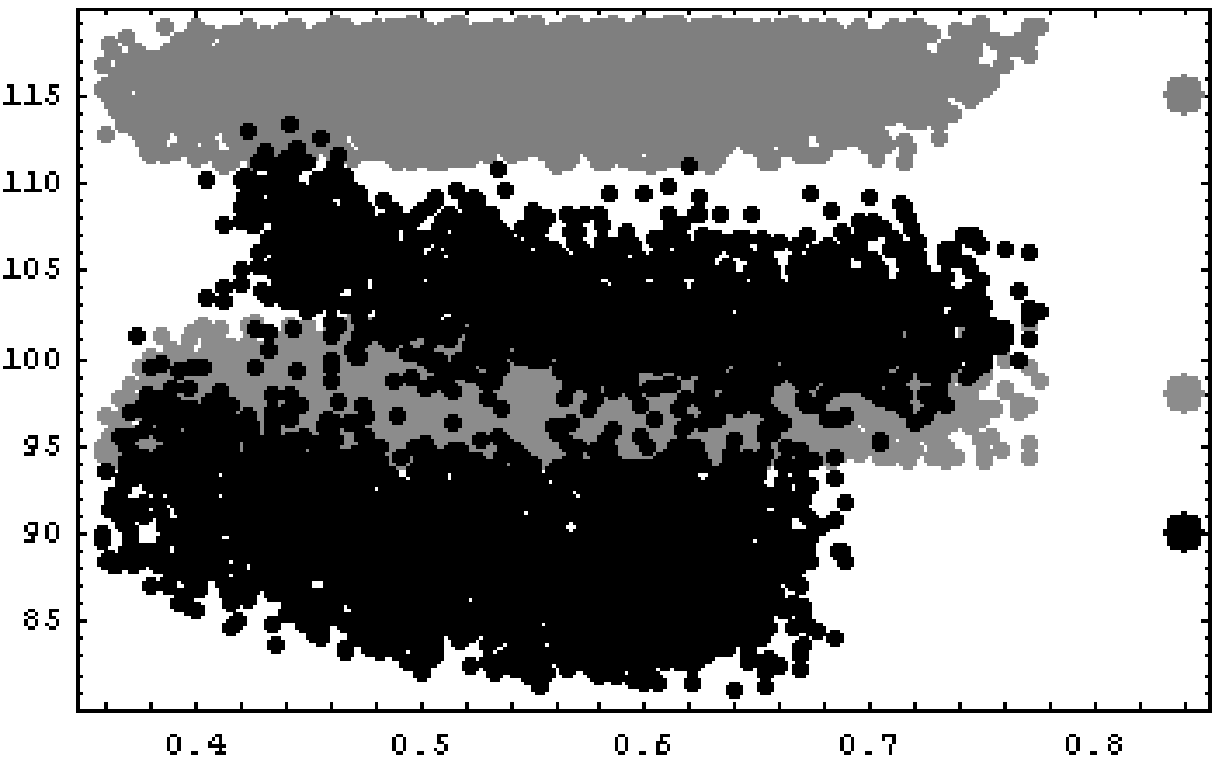}\vspace{.1cm}
\begin{picture}(0,0)(0,0)
   \put(-85,-15){$\mu_{eff}/v$}
     \put(-85,120){$h_s$}
     \put(-205,60){$M_{H}$}
     \put(-205,190){$M_{H}$}
     \put(-310,-15){$M_{Z^\prime}$}
     \put(-310,120){$A_t/v$}
    \put(-425,60){$M_{H}$}
    \put(-425,190){$M_{H}$}
  \put(-425,220){$(a)$}
    \put(-205,220){$(b)$}
    \put(-425,90){$(c)$}
    \put(-205,90){$(d)$}
\put(-425,180){\tiny{[GeV]}} \put(-425,50){\tiny{[GeV]}} \put(-205,180){\tiny{[GeV]}}
\put(-205,50){\tiny{[GeV]}}
     \put(-285,-14){\tiny{[GeV]}}
         \put(-26,51){\tiny{$m_h$}}
          \put(-26,95){\tiny{$m_H$}}
          \put(-26,30){\tiny{$m_A$}}
 \end{picture}
    \end{center}
    \caption{\label{c3m2}{The same as Fig.\ref{c3m1} but for Model II.}}
\end{figure}

Fig. \ref{cler2before} illustrates the impact of LEP bounds on the allowed parameter regions. Depicted are
variations of Higgs boson masses with $A_s/v$ in Model I (panels (a) and (c)) and Model II (panels (b) and (d)).
The Higgs boson masses in panels (a) and (b) are obtained only when the mass constraints $m_h \simeq 98\, {\rm
GeV}$ and $m_H \simeq 114\, {\rm GeV}$ are imposed. In these panels the pseudoscalar mass $m_A$ is seen to take
values in a rather wide range. What are shown in panels (c) and (d) are the allowed ranges of Higgs boson masses
when the constraint that the signal at $98\, {\rm GeV}$ forms only $\simeq 10 \%$ of the total
\cite{postlep,drees} is also included. This constraint, $C_{h Z Z}^2 \simeq 0.1$, is seen to have a significant
effect on the allowed ranges of $m_A$. Indeed, the allowed region for $m_A$ is seen to accumulate in mainly two
distinct domains: $m_A^{high} \sim m_H$ and $m_A^{low}\sim m_h$. This classification, however, is not precise at
all. First of all, $m_A^{high}$ and $m_A^{low}$ regions are not completely split; there are certain parameter
values for which this separation hardly makes sense. Next, in Model I, there are regions in the parameter space
where $m_{A}^{high}$ ($m_A^{low}$) lies visibly above $m_H$ ($m_h$). Finally, in Model II, $m_{A}^{high}$
($m_A^{low}$) lies significantly below $m_H$ ($m_h$) in almost entire parameter space. This figure is important
for revealing the impact of various constraints on the Higgs sector. In general, dynamical natures of $\mu$ and
$B$ parameters of the Higgs sector, their correlations with Z$^{\prime}$ mass, and their dependencies on various
model parameters (including the one-loop effects computed in \cite{everett}) result in certain differences from
the MSSM predictions \cite{kane,drees,plehn}. These reasons for these will be clear as we explore correlations
among the model parameters in LEP-allowed domains.

Continuing with Fig. \ref{cler2before}, one notes that the present
LEP data \cite{postlep} allow for $m_A$ to vary over a range that
covers both $m_h$ and $m_H$ such that, given the structure of the
allowed domains, there is a rough preference to either
$m_A^{low}\sim m_h$ or $m_A^{high}\sim m_H$. When $m_A \sim
m_A^{low}$ the pair-production process $e^+ e^- \rightarrow h A$
is kinematically allowed at LEP II energies. Moreover, since $C_{h
A Z}^2 \simeq C_{H Z Z}^2 \simeq 0.9$ the cross section does not
experience any significant suppression with respect to the SM
signal except for the fact that the overall signal is suppressed
with respect to $H Z$ production due to its p-wave nature.  The
separate LEP experiments have searched for associated $(h,H) + A$
production in the $b \overline{b} b \overline{b}$ and $b
\overline{b} \tau^+ \tau^-$ channels. However, a combined analysis
of the total LEP sample by all four collaborations is still not
available (except in preliminary form \cite{prel} which summarizes
the status as of summer 2005).

On the other hand, when $m_A \sim m_A^{high}$ the pair production process $e^+ e^- \rightarrow h A$ falls outside
the LEP II energy coverage. Moreover, besides p-wave suppression, number of such events should be a small
fraction of all such events since $C_{H A Z}^2 \simeq C_{h Z Z}^2 \simeq 10\, \%$. In either case, the present
LEP data favor pseudoscalar Higgs to have a mass roughly equaling $m_A^{high} \sim m_H$ or $m_A^{low}\sim m_h$.

\begin{figure}[htb]
\begin{center}
    \includegraphics[scale=.5]{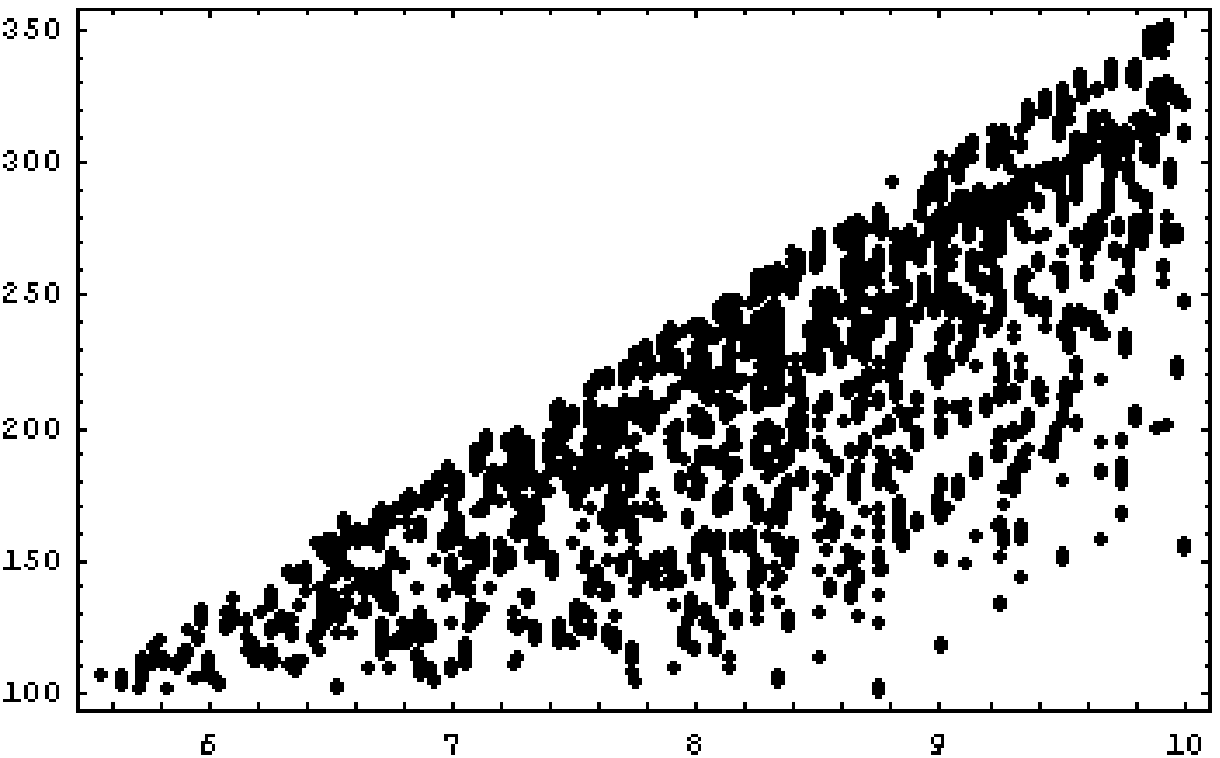}\hspace{1.5cm}
    \includegraphics[scale=.5]{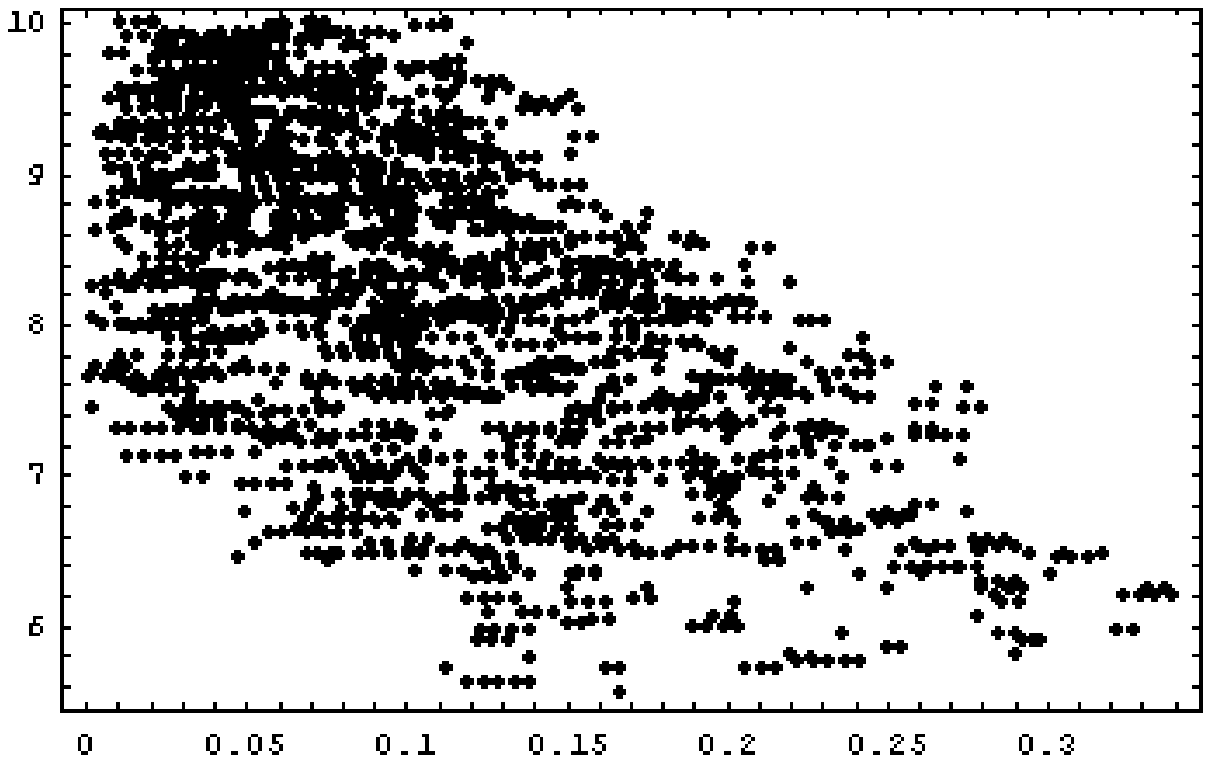}\vspace{1.cm}
    \includegraphics[scale=.5]{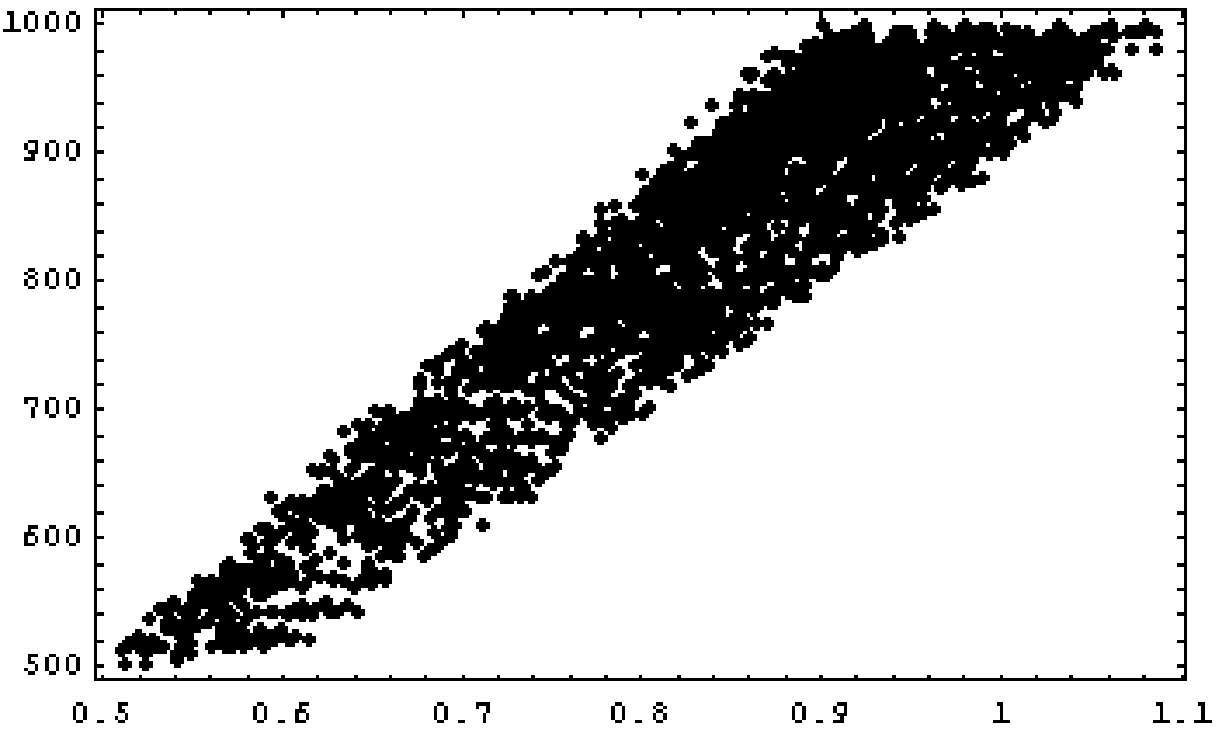}\hspace{1.5cm}
    \includegraphics[scale=.5]{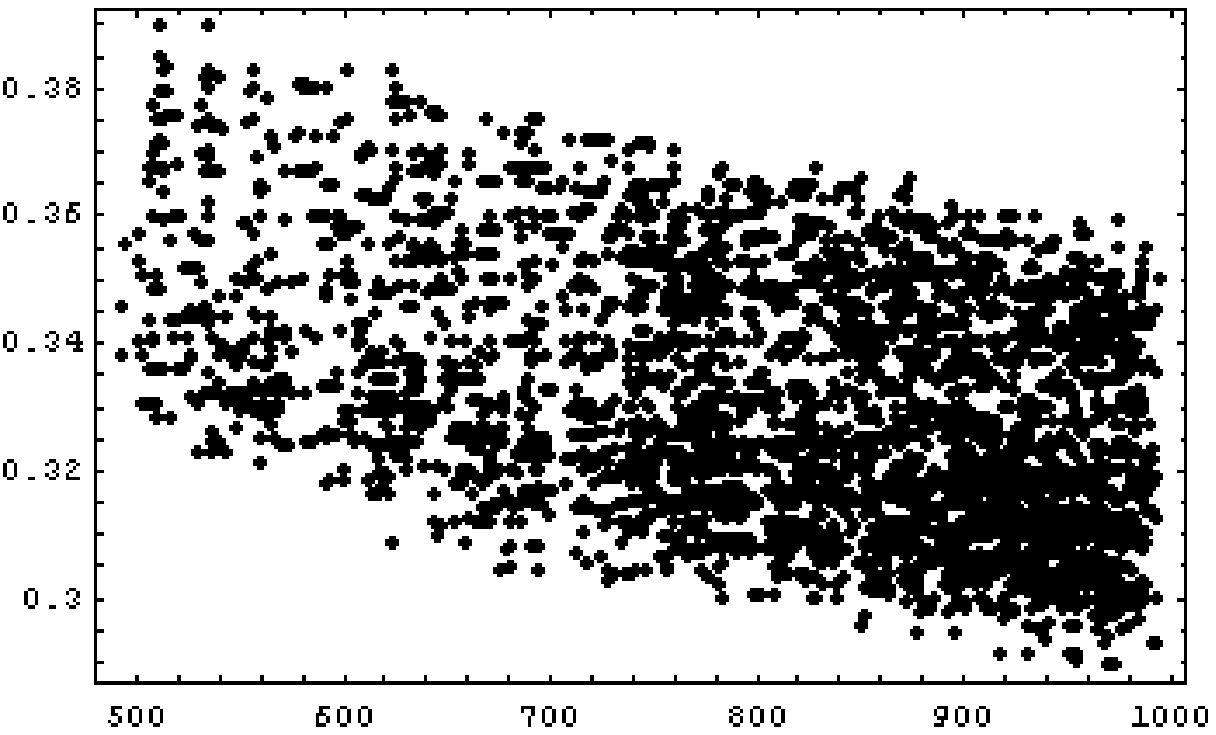}\hspace{0.cm}
\begin{picture}(0,0)(0,0)
     \put(-95,120){$A_s/v$}
    \put(-320,120){$A_t/v$}
     \put(-95,-15){$m_{H^\prime}$ \tiny{[GeV]}}
    \put(-320,-15){$\mu_{eff}/v$}
    \put(-425,185){$M_{\widetilde{t}_1}$}
    \put(-425,170){\tiny{[GeV]}}
    \put(-205,180){$A_t/v$}
    \put(-425,60){$M_{Z^\prime}$}
    \put(-205,60){$h_s$}
      \put(-425,45){\tiny{[GeV]}}

    \put(-425,230){$(a)$}
    \put(-205,230){$(b)$}
    \put(-425,90){$(c)$}
    \put(-205,90){$(d)$}
 \end{picture}
    \end{center}
    \caption{\label{cler3}{ Correlations among various model parameters. Bounds are similar for Model II. }}
\end{figure}

We now continue to explore correlations among various model parameters in light of the LEP bounds (\ref{bounds})
on the Higgs boson masses and couplings. Figs. \ref{c3m1} and \ref{c3m2} show how Higgs boson masses depend on
various parameters in Model I and Model II, respectively. These figures are particularly useful for determining
the allowed ranges of $h_s$ and $\mu_{eff}$ (see the panels (b) and (d) in each figure) while $M_{Z^{\prime}}$
varies in between the two limits and $A_t$ respects its upper bound (see panels (a) and (c) in each figure).
Depicted in Fig. \ref{cler3} are correlations among certain parameters in Model I. Furthermore, Table
\ref{table3} tabulates precise lower and upper (the numbers in front and inside the parentheses, respectively)
bounds on model parameters and resulting physical particle masses. These numbers are read off from the associated
data files. Below we provide a comparative analysis of various parameters illustrated in Figs. \ref{cler2before},
\ref{c3m1}, \ref{c3m2} and \ref{cler3} as well as the limits given in Table \ref{table3}:

\begin{itemize}
\item Low values of $A_t$ are disfavored. Indeed, $A_t \simgt 5\, v$ for bounds to be respected. Its minimal
value is determined by the lower bound imposed on the light stop mass, $m_{\widetilde{t}_1} \simgt 100\, {\rm
GeV}$ (see Fig. \ref{cler3} (a)). The precise ranges of $A_t$ for each model can be found in Table \ref{table3}.
In general, larger the $A_t$ smaller the $m_A$ because radiative correction to $m_A$ is proportional to $A_t$ and
it is negative at large $A_t$ where lighter stop weighs well below $M_{Z^{\prime}}$ \cite{everett}.

Moreover, the light stop mass varies with $A_t$ as in panel (a) of Fig. \ref{cler3}. The reason for this behavior
is that the soft masses $M_{\widetilde{Q}}$ and  $M_{\widetilde{U}}$ change in the background, and $A_t$ is
allowed to take larger values as their mean increases. As given in Table \ref{table3}, the light stop mass
remains below $\sim 360\, {\rm GeV}$ and heavy stop weighs above $\sim 660\, {\rm GeV}$. These masses are well
within the range which will be covered by searches at the LHC.

\item As suggested by Fig. \ref{cler2before} and panel (b) of Fig. \ref{cler3},  high values of $A_s$ are
disfavored. Indeed, $A_s$ is below  $v/3$ in both models (where
precise values can be found in Table \ref{table3}). The reason for
this is that at large $v_s$ (as needed to make Z$^{\prime}$ heavy
enough) $A_s$ is forced to take small values for making the
effective Higgs bilinear mixing $B_{eff}\propto h_s v_s A_s$ small
enough so that the two CP-even Higgs bosons (and necessarily the
pseudoscalar Higgs) weigh close to $M_Z$. In general, smaller the
$A_s$ lighter the $A$ boson (see panels (c) and (d) of Fig.
\ref{cler2before}) since $m_A^2 \propto B_{eff}$ at tree level. On
the other hand, radiative corrections are enhanced at large
$\mu_{eff} A_t$, and thus, $A_s$ takes small values at large $A_t$
to balance contributions of the one-loop corrections, as suggested
by the panel (b) of Fig. \ref{cler3} (see \cite{everett} for
dependencies of $m_A$ on various parameters).

\item As suggested by panels (c) of Figs. \ref{c3m1} and \ref{c3m2}, closer the $M_{Z^{\prime}}$ to its lower
bound larger the variation in pseudoscalar mass. This is expected since for light Z$^{\prime}$ the singlet VEV is
lowered and singlet compositions of $h$, $H$ and $A$ get pronounced. On the other hand, as $M_{Z^{\prime}}$ takes
on larger values, $m_A^{high}$ and $m_A^{low}$ domains allowed for $m_A$ tend to get closer to each other.
Therefore, the presence of two roughly distinct regions for $m_A$ is related to the extended nature of the Higgs
sector (or dynamical nature of the $\mu$ parameter). As it has already been reported in \cite{kane,drees,plehn},
the LEP bounds (\ref{bounds}) do not lead to such roughly split regions for $m_A$ in the MSSM. Note that gradual
decrease of the gap between $m_A^{high}$ and $m_{A}^{low}$ as  $M_{Z^{\prime}}$ increases is a signal of the
approach to MSSM limit. However, one keeps in mind that as $M_{Z^{\prime}}$ increases so does $\mu_{eff}$ unless
$h_s A_s$ is forced to take small values to keep doublet-dominated Higgs bosons light. This observation is
confirmed by panel (d) of Fig. \ref{cler3}. Though hard to confirm experimentally (since experiment will
eventually return a specific value for each Higgs boson mass), the aforementioned behavior of $m_A$ can be useful
for deciding on whether the model under concern is the MSSM or not. This can be accomplished if a certain set of
parameters $\mu_{eff}$, stop masses, soft parameters etc. is measured and their correlations are confronted with
predictions of the model.

\item The panels (b) and (d) of Figs. \ref{c3m1} and \ref{c3m2} as well as Table \ref{table3} reveal that $h_s$
and $\mu_{eff}/v$ are restricted to lie within narrow ranges below unity. That these parameters must be bounded
is clear from the upper bound on $m_h$ given in (\ref{mhupper}); for given values of U(1)$^{\prime}$ charges and
$g_Y^{\prime}$, the most $h_s$ can do is to vary within a certain interval in accord with the uncertainity in
$m_h$ value as well as radiative corrections. Indeed, $h_s \in \left[0.29, 0.32\right]$ in Model I and $h_s \in
\left[0.39, 0.43\right]$ in Model II. Similarly, $\mu_{eff} \in \left[0.36, 0.51\right] v$ in Model I and
$\mu_{eff} \in \left[0.77, 1.09\right] v$ in Model II. These restrictions arise from lightness of all
doublet-dominated Higgs bosons $h$, $H$ and $A$, and this is realized by rather small values of $h_s$. Indeed,
heavy Z$^{\prime}$ requires large values of $v_s$ with an indirect dependence on $h_s$ (see panels (c) and (d)
of Fig. \ref{cler3}) whereas the Higgs sector prefers small values of $h_s v_s$. These observations are further
supported by panels (c) and (d) of Fig. \ref{cler3}.

\item Table \ref{table3} depicts allowed ranges of the model parameters and corresponding predictions for Higgs
boson and stop masses when $M_{Z^{\prime}}$ varies in the ranges indicated. Scatter plots of some parameters in
this table are provided in Figs. \ref{cler2before}, \ref{c3m1}, \ref{c3m2} and \ref{cler3}. For each parameter,
the number in parenthesis shows the maximum value and the one in front the minimum value. As it should be clear
from the previous figures, some boundaries are already fixed with our choices ($e.g.$ larger values of $A_t$ is
possible but we keep it below $10 v$). In reading this table, it should be kept in mind that we have restricted
$M_{Z^{\prime}}$ into a rather conservative range. Indeed, once its upper bound is relaxed $\tan\beta$ will be
allowed to swing in a larger range (since then $|\alpha_{Z-Z^\prime}|$ allows for larger values of $\Delta^2$ as
illustrated in Fig.\ref{quqd0}), and it will lead to broadening of the allowed ranges of parameters. However,
even in this heavy Z$^{\prime}$ domain, the overall lightness of the Higgs sector will continue to bound $h_s$
and $A_s$ in ways similar to illustrations given in the figures.

\end{itemize}

\begin{table}[tbp]
\begin{center}
\begin{tabular}{|c||c|c|c||c|c|c|}\hline
 Inputs & Model I&Model II& Predictions (in GeV)& Model I& Model II\\ \hline \hline
$A_t/v$      &5.6 (10) & 5.3 (10)   & $M_{\widetilde{t}_1}$ & 101 (352)& 100 (365) \\\hline
$A_s/v$      &0 (0.34) & 0 (0.24)  & $M_{\widetilde{t}_2}$  & 665 (1130) & 658 (1202)\\\hline
$h_s$        & 0.29 (0.39) & 0.32 (0.43)  &$M_{Z^\prime}$  & 501 (1000) &502 (1000)\\\hline
$v_s/v$      & 2.1 (4.4) &1.4 (2.9)   &$m_h$    & 95 (101) &95 (101)  \\ \hline
$\mu_{eff}/v$ &0.51 (1.09)&0.36 (0.77) &$m_H$ &111 (119)&111 (119)  \\\hline
$M_{\widetilde{Q}}/v$  & 0.6 (4)& 1.2 (4.4) &$m_{H^\prime}$ &493 (995)&496 (996)\\\hline
$M_{\widetilde{U}}/v$  & 0.6 (4) & 0.6 (4.4) &$m_A$      & 86 (133)&81 (113) \\\hline
\end{tabular}
\end{center}
\caption{\label{table3}{Allowed ranges of input parameters and predictions for the particle masses in Model I and
Model II.}}
\end{table}

The analysis of the U(1)$^{\prime}$ parameter space presented in this section takes into account only the LEP
bounds (\ref{bounds}), and those resulting from the Z-Z$^{\prime}$ mixing. There exist, however, additional
indirect bounds from various observables like relic density of neutralinos \cite{barger1}, muon $g-2$
\cite{barger2} and rare processes \cite{barger3}. Normally, these additional constraints must also be taken into
account for a finer determination of the allowed parameter ranges (as has recently been performed by Hooper and
Plehn \cite{plehn} for the MSSM). In this work we have ignored bounds from such observables though this needs to
be confirmed by an explicit calculation.

\section{Conclusion}

In this work we have analyzed implications of LEP two-light-Higgs data on U(1)$^{\prime}$ models with
$M_{Z^{\prime}} \in \left[0.5, 1\right]\, {\rm TeV}$. We have depicted bounds on various parameters both by
scanning of the parameter space and by determining the maximal ranges of the individual parameters. Our results
suggest that the model is capable of reproducing the LEP results in wide regions of the parameter space.

We have found that, for CP-even Higgs bosons $h$ and $H$ to agree with the LEP data in masses and couplings,
($i$) the Higgs Yukawa coupling $h_s$ and the corresponding soft mass $A_s$ are forced to remain bounded in order
to keep the Higgs bosons under concern sufficiently light, ($ii$) the pseudoscalar Higgs boson weighs either
close to $m_h$ or $m_H$ with a finite gap in between. The bounded nature of these parameters stem from our
enhanced knowledge about the Higgs boson masses (according to LEP indications). The gap in between $m_A^{high}$
and $m_A^{low}$ bands tends to shrink with increasing $M_{Z^{\prime}}$.

The material presented in this work, in a more general setting, might be regarded as illustrating response of the
supersymmetric U(1)$^{\prime}$ models to constraints enforcing their Higgs sectors to be light. These models,
compared to MSSM, are known \cite{97makalesi,everett,05makalesi} to be capable of accommodating larger values for
the lightest Higgs boson mass already at tree level. Therefore, their potential to generate smaller values of the
lightest Higgs boson masses (as in, for instance, the LEP data \cite{postlep}) requires certain model parameters
to be restricted more strongly than in MSSM or NMSSM. In this sense, results reported in this work might serve as
a case study illustrating response of the $\mu$ problem solving models against constraints forcing their Higgs
sectors to weigh light.

\section{Acknowledgements}
The work of D. D. was partially supported by Turkish Academy of
Sciences through GEBIP grant, and by the Scientific and Technical
Research Council of Turkey through project 104T503. The work of L.
S. was partially supported by post-doctoral fellowship of the
Scientific and Technical Research Council of Turkey.

\end{document}